\documentclass[%
 aip,
 jcp,
 amsmath,amssymb,
 reprint,%
]{revtex4-1}

\usepackage{graphicx}
\usepackage{dcolumn}
\usepackage{physics}
\usepackage[utf8]{inputenc}
\usepackage[T1]{fontenc}
\usepackage{mathptmx}
\usepackage{etoolbox}
\usepackage{booktabs}
\usepackage{bbding}
\usepackage{hyperref}
\usepackage{textgreek}
\usepackage{float}
\usepackage{pifont}
\usepackage{siunitx}  
\usepackage{bm}

\newcommand{\ve}[1]{\bm{#1}}

\newcommand{\re}{\operatorname{Re}}
\newcommand{\im}{\operatorname{Im}}
\newcommand{\ii}{\mathrm{i}}
\newcommand{\cG}{\bm{\mathcal{G}}}
\newcommand{\cK}{\mathcal{K}}
\newcommand{\cP}{\mathcal{P}}
\newcommand{\eF}{\epsilon {\mathrm F}}
\newcommand{\Ct}{\tilde{\mathbf C}}
\newcommand{\D}{\mathbf D}
\newcommand{\Dn}{\mathbf D_n}
\newcommand{\Dm}[1]{\mathbf D_{m,#1}}
\newcommand{\Sm}{\mathbf S}

\makeatletter
\def\@email#1#2{%
 \endgroup
 \patchcmd{\titleblock@produce}
  {\frontmatter@RRAPformat}
  {\frontmatter@RRAPformat{\produce@RRAP{*#1\href{mailto:#2}{#2}}}\frontmatter@RRAPformat}
  {}{}
}%
\makeatother
\begin{document}


\title{Relativistic Kohn--Sham Theory of Paramagnetic Magnetic-Resonance Properties}

\author{Christof Holzer$^*$}
\affiliation{Institute of Quantum Materials and Technologies, Karlsruhe
Institute of Technology (KIT), Kaiserstra\ss{}e 12, 76131 Karlsruhe,
Germany}
\email{christof.holzer@kit.edu}

\author{Yannick J. Franzke$^*$}
\affiliation{Institute of Nanotechnology, Karlsruhe
Institute of Technology (KIT), Kaiserstra\ss{}e 12, 76131 Karlsruhe, Germany}
\email{yannick.franzke@kit.edu}

\begin{abstract}
A general density functional theory based relativistic theory for 
paramagnetic NMR (pNMR) shifts, EPR g-tensors and hyperfine coupling 
tensors as well as magnetizabilities is derived. The formulation 
uses spin--orbit exact two-component (X2C) theory,
gauge-including atomic orbitals, and generalized complex Kohn--Sham
response equations with all ingredients being implemented analytically
with the help of Goldstone modes and an orientation-constrained shielding.
Additionally, we describe a two-component augmented Roothaan--Hall (ARH)
solver for robust and stable convergence, which ensures that both a 
stationary point and a minimum are found. This is crucial for the convergence 
of the response equations for properties of paramagnetic systems. The extension of
common population analyses for the two-component theory to aid in the
interpretation of the spectra is further described. Using these new techniques, 
we are able to analyze the success and failure of density functional theory 
for relativistic open-shell systems including actinide complexes.
\end{abstract}

\maketitle

\section{Introduction}
\label{sec:introduction}
Open-shell species with unpaired electrons and non-zero electron spin such
as organic radicals or transition-metal as well as lanthanide or actinide
complexes are of great importance for chemical reactions or quantum materials
and often reveal uncommon or novel physical properties and electronic structures.
The given species are commonly studied with paramagnetic nuclear resonance
(pNMR) or electron paramagnetic resonance (EPR) spectroscopy.
EPR and pNMR properties are not only of relevance for analytics in catalysis
and single molecule magnets in fundamental research~\cite{Bertini.Luchinat.ea:2001,
Kaupp.Kohler:Combining.2009, Pell.Pintacuda.ea:Paramagnetic.2019, Goldfarb.Stoll:2018,
Moon.Cornella:Bismuth.2022, Pang.Nothling.ea:Synthesis.2023}
but are of great technological importance, as demonstrated by the recent work
in quantum information science or quantum devices utilizing quantum-coherent
control. \cite{Shiddiq.Komijani.ea:Enhancing.2016, Kundu.White.ea:Clock.2022,
Gimenez-Santamarina.Cardona-Serra.ea:Exploiting.2020, Ngo.McClain.ea:Large.2025,
Willke.Bae.ea:Hyperfine.2018} Typically, the considered chemical systems involve
(heavy) metal atoms and relativistic effects become important for the paramagnetic
resonance phenomena. \cite{InB-Neese:2017, Autschbach:Relativistic.2014,
Novotny.Komorovsky.ea:Paramagnetic.2024}
However, even with significant theoretical progress over the last 25 years,
\cite{Rinkevicius.Vaara.ea:Calculations.2003, InB-Moon.Patchkovskii:Calculation.2004, 
Pennanen.Vaara:Nuclear.2008, Liimatainen.Pennanen.ea:1H.2009,
Autschbach.Patchkovskii.ea:Calculation.2011, Van-den-Heuvel.Soncini:NMR.2012,
Van-den-Heuvel.Soncini:NMR.2013, Soncini.Van-den-Heuvel:Communication.2013, 
Komorovsky.Repisky.ea:Four-Component.2013, Gendron.Sharkas.ea:Calculating.2015,
Martin.Autschbach:Temperature.2015, Martin.Autschbach:Erratum.2016,
Vaara.Rouf.ea:Magnetic.2015, Rouf.Mares.ea:1H.2015, Rouf.Mares.ea:Relativistic.2017, 
Lang.Ravera.ea:Solution.2020, Pyykkonen.Vaara:Computational.2023, Franzke.Holzer:Impact.2022,
Franzke.Bruder.ea:Paramagnetic.2024, Pyykkonen.Wodynski.ea:First-principles.2025,
InB-Islam.Bolvin:2025}
the computation of pNMR properties remains very challenging and often
involves the \textit{ad hoc} combination of various theoretical frameworks
and relativistic methodologies. Among the various challenges are the
accurate description of the open-shell ground-state and its proper
self-consistent field (SCF) convergence, considering both static and
dynamic correlation to a sufficient degree, the thorough treatment of
scalar-relativistic effects and spin--orbit coupling (SOC), as well
as the inclusion of solvation effects or even molecular dynamics.
Compared to the related calculation of EPR parameters such as the
hyperfine coupling constant and the g-tensor of transition-metal
and p-block elements, \cite{Gohr.Hrobarik.ea:Four-Component.2015,
Franzke.Yu:Hyperfine.2022, Franzke.Yu:Quasi-Relativistic.2022,
Feng.Duignan.ea:Electron-Nucleus.2021, Saitow.Neese:Accurate.2018,
Windom.Perera.ea:Benchmarking.2022, Glasbrenner.Vogler.ea:Gauge-origin.2018}
accurate pNMR shifts are often even more difficult to obtain, as the
spin-excess density needs to be correctly described not only at the
paramagnetic center but over the whole molecule.

In this work, we derive a general relativistic Kohn--Sham density functional
theory (DFT) of pNMR and EPR properties in a non-collinear exact two-component
\cite{Ilias.Saue:infinite-order.2007, Kutzelnigg.Liu:Quasirelativistic.2005,
Surjuse.Valeev:SAP-X2C.2026} (X2C) framework.
This approach treats scalar-relativistic and spin--orbit effects on an equal
footing. All parameters are available by solving general complex
response equations and EPR parameters are available from one SCF run
only---in contrast to the common 3-SCF approach.
\cite{Gohr.Hrobarik.ea:Four-Component.2015, Franzke.Yu:Hyperfine.2022,
Franzke.Yu:Quasi-Relativistic.2022, Verma.Autschbach:Relativistic.2013,
Verma.Autschbach:Variational.2013, Malkin.Malkina.ea:Relativistic.2005,
Misenkova.Lemken.ea:four-component.2022}
The key of the theoretical formulation is a thorough analysis of
spin-orientation symmetry breaking, orientation-constrained orbital
shielding, and the definition of so-called Goldstone modes which are
relaxed tangents and allow to extract the pNMR/EPR properties.

\section{Theory}
\label{sec:2c-nonkramers-nmr}

\subsection{Preliminaries}
We consider a spin-polarized, complex two-component (2c) odd-electron Kohn--Sham
(KS) determinant as one member of an isolated Kramers doublet generated by time
reversal. The effective pseudospin is $\widetilde S=1/2$ throughout. Its
relation to the physical spin is specified in Sec.~\ref{subsec:kramers-spin-map}.
The one-particle density is the Hermitian matrix formed from the KS
spinors $|\phi_i\rangle$ and their occupations $n_i$ as
\begin{equation}
  \mathbf D= \sum_i n_i\,|\phi_i\rangle\langle\phi_i| = 
  \begin{pmatrix}
    \mathbf D^{\alpha\alpha} & \mathbf D^{\alpha\beta}\\
    \mathbf D^{\beta\alpha}  & \mathbf D^{\beta\beta}
  \end{pmatrix}
  \label{eq:2c-density}
\end{equation}
and all eight one-component (1c) real symmetric/antisymmetric atomic-orbital (AO) density
channels associated with its four complex spin blocks are retained. That is, the 2c density
matrix describes the total particle density and paramagnetic current density as
well as their spin counterparts with the symmetric/antisymmetric real/imaginary
linear combinations. In the Pauli form, the density matrix is
partitioned into
\begin{equation}
  \mathbf D =\frac12\left(\,\mathbf{D}_n I +{\mathbf{D}_m}\cdot\ve\sigma\right), \qquad {\mathbf{D}_m}\cdot\ve\sigma=\sum_{k=x,y,z}{\mathbf{D}}_{m, k}\sigma_k
  \label{eq:pauli}
\end{equation}
where $I$ is the $2\times2$ identity
\begin{equation}
  I=\begin{pmatrix}1&0\\0&1\end{pmatrix}
\end{equation}
and
$\ve\sigma=(\sigma_x,\sigma_y,\sigma_z)$ is the vector of the Pauli spin matrices,
\begin{equation}
  \sigma_x=\begin{pmatrix}0&1\\1&0\end{pmatrix},
  \sigma_y=\begin{pmatrix}0&-\text{i}\\\text{i}&0\end{pmatrix},
  \sigma_z=\begin{pmatrix}1&0\\0&-1\end{pmatrix}.
\end{equation}
The indices $i, j, \dots $ denote occupied spinors, $a, b, \dots$ are virtual spinors,
and $p, q, \dots $ refer to general spinors. Greek letters denote the AO
basis functions. 

For a nucleus $N$, the observable pNMR shielding
is separated into an orbital and a spin-manifold paramagnetic contribution,
\begin{equation}
  \boldsymbol\sigma_N(T)
  =\boldsymbol\sigma_N^{\mathrm{orb}}
  +\boldsymbol\sigma_N^{\mathrm{para}}(T).
  \label{eq:shielding-split}
\end{equation}
Only the latter depends explicitly on the temperature $T$.

\subsection{Symmetry Breaking and its Consequences}
In the absence of spin--orbit coupling (SOC), the electronic Hamiltonian is
invariant under global $\mathrm{SU}(2)$ spin rotations. At the mean-field
level, the locally collinear determinant underlying the present construction,
with $\langle\hat{\mathbf S}\rangle\ne\mathbf0$, selects one magnetization
direction and spontaneously realizes the breaking pattern
\begin{equation}
  \mathrm{SU}(2)\longrightarrow\mathrm{U}(1),
  \label{eq:su2-breaking}
\end{equation}
where the residual $\mathrm{U}(1)$ consists of rotations about the selected
axis. \cite{Kasper.Jenkins.ea:Perspective.2020}
This mean-field symmetry breaking yields two transverse Goldstone coordinates
on $\mathrm{SU}(2)/\mathrm{U}(1)\simeq S^2$. It does not imply symmetry
breaking of the exact many-body ground state. We define
\begin{equation}
  \mathbf n=\frac{\langle\hat{\mathbf S}\rangle}
                  {|\langle\hat{\mathbf S}\rangle|},
  \qquad
  \mathbf e_1\perp\mathbf n,
  \quad \mathbf e_2\perp\mathbf n,
  \quad \mathbf e_1\perp\mathbf e_2,
\end{equation}
with $(\mathbf e_1,\mathbf e_2,\mathbf n)$ right handed. If SOC is absent,
transverse rotations are zero modes of the SCF orbital Hessian, while axial
rotation changes only the phase of a collinear determinant. SOC lifts the
transverse modes to finite-curvature pseudo-Goldstone directions.

We use three collective coordinates throughout. For rigid spin rotations,
let $\boldsymbol\vartheta=(\vartheta_1,\vartheta_2,\vartheta_3)^{\mathrm T}$
denote angles in radians and define
\begin{equation}
  |\Phi_{\mathrm{rigid}}(\boldsymbol\vartheta)\rangle
  =\exp\!\left[-\mathrm i
    (\vartheta_1\hat S_{\mathbf e_2}
     -\vartheta_2\hat S_{\mathbf e_1}
     +\vartheta_3\hat S_{\mathbf n})\right]|\Phi_0\rangle .
  \label{eq:three-spin-coordinates}
\end{equation}
The first two angles tilt the spin axis toward $\mathbf e_1$ and
$\mathbf e_2$, respectively. Rotation about $\mathbf n$ leaves the total
spin unchanged to first order,
\begin{equation}
  \delta_3\langle\hat{\mathbf S}\rangle
  =\vartheta_3\,\mathbf n\times\langle\hat{\mathbf S}\rangle
  =\mathbf0,
  \label{eq:third-collective-tangent}
\end{equation}
but can change a non-collinear determinant. This third coordinate is
redundant only when the occupied--virtual block of $\hat s_{\mathbf n}$
vanishes, indicating a continuous residual $\mathrm{U}(1)$. We consider
the fully broken case in which all three generators are independent.

Let $\boldsymbol{\zeta}$ denote the real representation of the complex
occupied--virtual orbital rotations. For an axis $\mathbf u$, the rigid
spin generator and its occupied--virtual block are
\begin{equation}
  \hat s_{\mathbf u}=\frac12\mathbf u\!\cdot\!\boldsymbol\sigma,
  \qquad
  \kappa^{(\mathbf u)}_{ai}=-\mathrm i
  \langle\phi_a|\hat s_{\mathbf u}|\phi_i\rangle,
  \qquad
  \kappa^{(\mathbf u)}_{ia}=-\kappa^{(\mathbf u)*}_{ai}.
  \label{eq:goldstone-generator}
\end{equation}
The AO representation of $\hat s_{\mathbf u}$ contains the spatial overlap
matrix. With $M^{(\mathbf u)}_{ai}=\langle\phi_a|
\hat s_{\mathbf u}|\phi_i\rangle$, define the real rigid tangent
\begin{equation}
  \mathbf r(\mathbf u)=
  \begin{pmatrix}
    \operatorname{Re}\boldsymbol\kappa^{(\mathbf u)}\\
    \operatorname{Im}\boldsymbol\kappa^{(\mathbf u)}
  \end{pmatrix}
  =\begin{pmatrix}
    \operatorname{Im}\mathbf M^{(\mathbf u)}\\
    -\operatorname{Re}\mathbf M^{(\mathbf u)}
  \end{pmatrix},
  \label{eq:goldstone-real-vector}
\end{equation}
and set
\begin{equation}
  \mathbf r_1=\mathbf r(\mathbf e_2),\qquad
  \mathbf r_2=\mathbf r(-\mathbf e_1),\qquad
  \mathbf r_3=\mathbf r(\mathbf n).
  \label{eq:rigid-tangents}
\end{equation}
Without SOC, the transverse rigid tangents are Hessian zero modes
but SOC mixes them with orbital relaxation.
\cite{Bauernschmitt.Ahlrichs:Stability.1996}

We normalize the transverse production coordinates to changes of the
measured physical spin. Let $s_0=|\langle\hat{\mathbf S}\rangle|$ and
$\mathbf P=(\mathbf p_1,\mathbf p_2)$ contain the raw orbital gradients of
$\mathbf e_\mu\!\cdot\!\langle\hat{\mathbf S}\rangle$ in the response
convention used below. Their rigid Jacobian is
\begin{equation}
  \Xi_{\mu\nu}=\mathbf p_\mu^{\mathrm T}\mathbf r_\nu
  =\operatorname{Im}\operatorname{Tr}_{\rm ov}
  \left(\mathbf M^{(\mathbf e_\mu)\dagger}
              \mathbf M^{(\mathbf u_\nu)}\right),
  \quad \mu,\nu\in\{1,2\},
  \label{eq:orientation-covectors}
\end{equation}
where $\mathbf u_1=\mathbf e_2$ and $\mathbf u_2=-\mathbf e_1$.
Resolving the virtual projector as $\mathbf1-\mathbf P_{\rm occ}$ leaves
an occupied--occupied trace that is real and drops out. The identity
$\hat s_{\mathbf a}\hat s_{\mathbf b}
=\tfrac14(\mathbf a\!\cdot\!\mathbf b)
+\tfrac{\mathrm i}{2}\hat s_{\mathbf a\times\mathbf b}$ then gives
\begin{equation}
  \Xi_{\mu\nu}
  =\tfrac12\langle\hat S_{\mathbf e_\mu\times\mathbf u_\nu}\rangle
  =\frac{s_0}{2}\delta_{\mu\nu}.
  \label{eq:analytic-orientation-jacobian}
\end{equation}
Consequently, the physical-spin covectors are simply
\begin{equation}
  \mathbf J_\mu=2\mathbf p_\mu,\qquad \mu=1,2.
  \label{eq:analytic-covectors}
\end{equation}
The numerical evaluation of $\boldsymbol{\Xi}$ serves as a convention check.

The rank-two Watanabe--Brauner form
$\rho_{ab}=\epsilon_{abc}\langle\hat S_c\rangle$ has kernel
$\mathbf n$, so
$\mathbf p_\mu^{\mathrm T}\mathbf r_3
=\tfrac12\langle\hat S_{\mathbf e_\mu\times\mathbf n}\rangle=0$.
Total-spin observables therefore cannot normalize the third coordinate.
We use the real part of the quantum geometric tensor, 
which reduces to the Euclidean product in orthonormal
occupied--virtual coordinates. Taking $\mathbf r_3$ as its own metric dual
and deflating against the transverse pair gives
\begin{equation}
\begin{aligned}
  \mathbf J_3=\frac{1}{\Xi_{33}}
  \left(\mathbf r_3-
    \frac{1}{s_0}\sum_{\mu=1,2}\Xi_{3\mu}\mathbf J_\mu\right),\\
  \Xi_{3\mu}=\mathbf r_3^{\mathrm T}\mathbf r_\mu,\quad
  \Xi_{33}=\|\mathbf r_3\|^2>0.
\end{aligned}
  \label{eq:third-covector}
\end{equation}
The coordinate vector $\mathbf q=(q_1,q_2,q_3)^{\mathrm T}$ thus contains
two physical-spin displacements and one angle in radians. Its normalized
rigid tangents and covectors satisfy
\begin{equation}
  \mathbf Q=\left(\frac{\mathbf r_1}{s_0},
                 \frac{\mathbf r_2}{s_0},\mathbf r_3\right),
  \qquad
  \mathbf J=(\mathbf J_1,\mathbf J_2,\mathbf J_3),
  \qquad \mathbf J^{\mathrm T}\mathbf Q=\mathbf1_3.
  \label{eq:full-coords}
\end{equation}
For rigid rotations, $q_\mu=s_0\vartheta_\mu$ ($\mu=1,2$) and
$q_3=\vartheta_3$ to first order. The relaxed SCF manifold is parameterized
in the same production coordinates as
\begin{equation}
\begin{aligned}
  \boldsymbol\zeta(\mathbf q)
  = & \mathbf T\mathbf q
   +\frac12\sum_{a,b=1}^{3}\mathbf z_{ab}q_aq_b
   +\mathcal O(q^3),\\
  \mathbf J^{\mathrm T}\boldsymbol\zeta(\mathbf q)
  = & \mathbf q+\mathcal O(q^3),
\end{aligned}
  \label{eq:relaxed-coords}
\end{equation}
where $\mathbf T$ and $\mathbf z_{ab}$ are the first-order and second-order
relaxed responses defined below.

Overall, we solve the coupled-perturbed Kohn--Sham (CPKS) equations with the 2c Hessian
$\mathbf H$,\cite{Pausch.Holzer:Linear.2022}
\begin{align}
  \mathbf H\mathbf Y= & \mathbf J, \\
  \mathbf R= & \mathbf J^{\mathrm T}\mathbf Y
           =\mathbf J^{\mathrm T}\mathbf H^{-1}\mathbf J,
  \label{eq:orientation-compliance}
\end{align}
where Hessian reciprocity requires $\mathbf R=\mathbf R^{\mathrm T}$. For a
non-singular $\mathbf R$, the relaxed orientation tangents are
\begin{equation}
  \mathbf T=\mathbf H^{-1}\mathbf J
  (\mathbf J^{\mathrm T}\mathbf H^{-1}\mathbf J)^{-1}
  =\mathbf Y\mathbf R^{-1},
  \qquad \mathbf J^{\mathrm T}\mathbf T=\mathbf1_3.
  \label{eq:relaxed-tangents}
\end{equation}
The matrix $\mathbf R$ is therefore the relaxed orientation compliance, and
\begin{equation}
  \mathbf K=\mathbf R^{-1}=\mathbf T^{\mathrm T}\mathbf H\mathbf T
  \label{eq:orientation-stiffness}
\end{equation}
is the $3\times3$ orientation stiffness. Its columns are the Lagrange
multipliers in the Karush--Kuhn--Tucker (KKT) system
$\mathbf H\boldsymbol\zeta_a=\mathbf J\boldsymbol\lambda_a$,
$\mathbf J^{\mathrm T}\boldsymbol\zeta_a=\boldsymbol\epsilon_a$.
The quadratic energy correction is 
$E^{(2)}=\tfrac12\mathbf q^{\mathrm T}\mathbf K\mathbf q$.
We refer to the relaxed SOC-lifted tangents as ``Goldstone modes'', 
given their relation to the transverse zero modes of the SOC-free limit.

Embedding tangent $\mathbf t_a^{\mathrm{orb}}$ in the anti-Hermitian matrix
$\boldsymbol\kappa_a$ leads to the first-order density,
\begin{equation}
  \mathbf D(q_a)=
  e^{q_a\boldsymbol\kappa_a}\mathbf D_0
  e^{-q_a\boldsymbol\kappa_a},
  \qquad
  \dot{\mathbf D}_a=
  \left[\boldsymbol\kappa_a,\mathbf D_0\right]
  \label{eq:tangent-density}
\end{equation}
with $\mathbf D_0$ denoting the ground-state density matrix.
All four complex spin blocks of $\dot{\mathbf D}_a$ are used in the property
contractions.

\subsection{Orientation-Constrained Orbital Shielding}

The orbital shielding is the mixed derivative with respect to the external
field component $B_p$ and the nuclear magnetic moment component $m_{Nq}$,
\begin{equation}
  \sigma^{\mathrm{orb}}_{N,pq}
  =\left.
   \frac{\mathrm d^2 E}{\mathrm dB_p\,\mathrm dm_{Nq}}
   \right|_{0,\,\mathbf J^{\mathrm T}\boldsymbol\zeta=0},
  \label{eq:orbital-shielding-definition}
\end{equation}
where the constraint excludes first-order motion along any of the collective
spin-orientation coordinates. Gauge-including atomic orbitals (GIAOs),
\begin{equation}
  \chi_\mu(\mathbf B)=
  \exp\!\left[-\frac{\mathrm i}{2c}
  (\mathbf B\!\times\!\mathbf R_\mu)\!\cdot\!\mathbf r\right]\chi_\mu,
  \label{eq:giao}
\end{equation}
are used for the explicit integral, overlap, and X2C picture-change
derivatives. \cite{Yoshizawa.Hada:Calculations.2017, Franzke.Holzer:Exact.2023, InB-Franzke:2025}
$c$ denotes the speed of light, $\chi$ a real-valued atomic orbital
basis function centered at a nucleus $\mu$, $\mathbf{R}$ and $\mathbf{r}$
are the nuclear and electronic coordinate vectors.

Let $\mathbf b^{P}=\partial^2E/(\partial\boldsymbol\kappa\,\partial P)$
be the orbital gradient generated by perturbation $P$. The stationarity
condition leads to the CPKS equation
\begin{equation}
  \mathbf H\mathbf x^{P}=-\mathbf b^{P}.
  \label{eq:cphf}
\end{equation}
To avoid counting spin reorientation in both shielding contributions, 
we use the KKT-constrained orbital response
\begin{equation}
  \begin{pmatrix}
    \mathbf H & \mathbf J\\
    \mathbf J^{\mathrm T} & \mathbf0
  \end{pmatrix}
  \begin{pmatrix}\mathbf x^P_\perp\\\boldsymbol\eta\end{pmatrix}
  =-
  \begin{pmatrix}\mathbf b^P\\\mathbf0\end{pmatrix},
  \qquad \mathbf J^{\mathrm T}\mathbf x^P_\perp=\mathbf0,
  \label{eq:constrained-response}
\end{equation}
where $\eta$ are the Lagrange multipliers.
Using the unconstrained solution
$\mathbf x^P=-\mathbf H^{-1}\mathbf b^P$ and
Eq.~\eqref{eq:relaxed-tangents}, the Schur correction is simply
\begin{equation}
  \mathbf x^P_\perp
  =\mathbf x^P-\mathbf T\mathbf J^{\mathrm T}\mathbf x^P.
  \label{eq:schur-projection}
\end{equation}

As in the closed-shell limit, \cite{Franzke.Holzer:Exact.2023,
Yoshizawa.Hada:Calculations.2017, Komorovsky.Repisky.ea:Fully.2010,
Olejniczak.Bast.ea:simple.2012, Xiao.Sun.ea:Fully.2012, Autschbach:role.2013}
the GIAO first-order density also contains an occupied--occupied metric
term from the field derivative of the AO overlap, which is absent from
the occupied--virtual response vector.
The constrained mixed derivative is
\begin{equation}
  \sigma^{\mathrm{orb}}_{N,pq}
  =E^{B_p m_{Nq}}_{\mathrm{explicit}}
   +(\mathbf b^{m_{Nq}})^{\mathrm T}\mathbf x^{B_p}_\perp
  =E^{B_p m_{Nq}}_{\mathrm{explicit}}
   -(\mathbf b^{m_{Nq}})^{\mathrm T}
    \mathbf H^{-1}_{\perp}\mathbf b^{B_p},
  \label{eq:orbital-shielding-response}
\end{equation}
where
$\mathbf H^{-1}_{\perp}=(\mathbf1-\mathbf T\mathbf J^{\mathrm T})
\mathbf H^{-1}=\mathbf H^{-1}-\mathbf H^{-1}\mathbf J\mathbf R^{-1}
\mathbf J^{\mathrm T}\mathbf H^{-1}$ is the symmetric KKT-constrained inverse.

\subsection{Right-Hand Side of the CPKS Equations}
\label{subsec:rhs}
The right-hand side of the CPKS equations for the magnetic field {$\mathbf B$} consists of the
full 2c GIAO derivative terms, including the one-electron X2C Hamiltonian, \cite{Franzke.Holzer:Exact.2023}
the overlap integrals, Coulomb, exchange, and exchange-correlation potential as well
as the occupied-occupied response terms of the Coulomb, exact exchange, and exchange-correlation
kernels. For the latter, this also involves the GIAO response to the density variables
\cite{Franzke.Holzer:Exact.2023, Olejniczak.Bast.ea:simple.2012, Komorovsky.Repisky.ea:Fully.2010}
\begin{equation}
    \mathbf{b}^B_{ai} = \mathbf{b}^{B, \text{1e}}_{ai} + \mathbf{b}^{B, \text{2e}}_{ai} =
    \left(\mathbf{h}^B_{ai} - {\bf\epsilon}_i \mathbf{S}^B_{ai} \right) +
    \left(\mathbf{b}^{B, {\text{pot}}}_{ai} + \mathbf{b}^{B, \text{kernel}}_{ai} \right) \text{.}
\end{equation}
The one-electron terms are unchanged from the closed-shell limit and were previously
described in detail. \cite{Franzke.Holzer:Exact.2023} Likewise, the extension of the
exact exchange integrals from closed-shell to open-shell for $\mathbf{b}^{{\text{pot}}}$
is straightforward as this simply requires to use all eight linear combinations of for the 2c
density matrix.
Overall, this leads to four symmetric and four antisymmetric terms, while the closed-shell
limit only has one antisymmetric part from the particle density but three symmetric contributions
from the spin-current densities. The GIAO crossproducts are antisymmetric and imaginary and
thus flip the symmetry. The Coulomb potential derivative requires a major extension compare
to the closed-shell limit. We use the multipole-accelerated resolution of the identity approximation
to the Coulomb terms (MARI-J/RI-J). The derivative of the Coulomb potential in RI-J leads to well
known the imaginary antisymmetric integrals due to the GIAO crossproduct term.
\cite{Reiter.Mack.ea:Calculation.2018, Loibl.Schutz:NMR.2012, Stoychev.Auer.ea:Self-Consistent.2018}
However, in 2c pNMR also real symmetric integrals associated with the RI-J auxiliary functions
$P$, $Q$ and their fitting coefficients $c_P$, $c_Q$ arise according to
\begin{equation}
  \frac{\partial J^{\text{RI-J}}_{\mu\nu}}{\partial B_p}
  = {\sum_{P}\frac{\partial (\mu\nu|P)}{\partial B_p}\,c_{P}}
  + {\sum_{P} (\mu\nu|P)\,\frac{\partial c_{P}}{\partial B_p}}
  \label{eq:RI-J-B}
\end{equation}
with the imaginary antisymmetric integral derivatives $\partial(\mu \nu|P)/\partial B_p$.
Here, the Mulliken notation is applied for the atomic-orbital space. 
We note that derivatives
are formed in the limit of a vanishing magnetic field as usual. 
The transformation of the integrals
into the occupied-virtual spinor space is straightforward.
The RI-J coefficient response obtained from the derivative of the fitting
equations,
\begin{align}
  \frac{\partial c_{P}}{\partial B_p}
     = & \sum_{Q} \left[(P|Q)^{-1}\right]_{PQ}\, \frac{\partial \gamma_{Q}}{\partial B_p}
     \label{eq:rij-coeffB} \\
  \frac{\partial \gamma_{Q}}{\partial B_p}
    = & \sum_{\lambda\sigma} \frac{\partial (\lambda\sigma|Q)}{\partial B_p}\,
       D_{\text{tot},\sigma\lambda}^{\mathrm{IA}} \textrm{.}
  \label{eq:gammaprime}
\end{align}
The term only arises through the imaginary antisymmetric particle density contribution,
$\textbf{D}^{\mathrm{IA}}_{\text{tot}} = \text{Im}\big(\textbf{D}^{\alpha \alpha}\big) + \text{Im}\big(\textbf{D}^{\beta \beta}\big)$,
which is associated with the particle current density of the ground state
induced by SOC. Therefore, the latter term in Eq.~\eqref{eq:RI-J-B} vanishes
in a Kramers-restricted formalism \cite{Franzke.Holzer:Exact.2023} or
if SOC is neglected. For properties which directly follow as energy derivatives
without response terms such as the g-tensor \cite{Franzke.Yu:Quasi-Relativistic.2022}
this new term is not needed. It is only needed when the given integrals
once meet a density other than the one from the SCF procedure. The derivative
of the exchange-correlation potential follows as
\begin{equation}
\begin{split}
    &  \bm{b}^{B_p, \text{pot, XC}}_{\mu \nu, u}  =
     \frac{\text{i}}{2c} \int v^{\text{XC}}_{\text{LSDA}, u} (\mathbf{R}_{\mu \nu} \crossproduct \mathbf{r})_p ~ \chi_{\mu} \chi_{\nu} ~ d\mathbf{r} \\
     & + \frac{\text{i}}{2c} \int v^{\text{XC}}_{\text{GGA}, u} \boldsymbol{\nabla}\rho(\mathbf{r}) ~ (\mathbf{R}_{\mu \nu} \crossproduct \mathbf{r})_p 
    \left[ \boldsymbol{\nabla} \chi_{\mu} \chi_{\nu} +  \chi_{\mu} \boldsymbol{\nabla} \chi_{\nu} \right]  d\mathbf{r} \\
    & +  \frac{\text{i}}{2c} \int\ v^{\text{XC}}_{\text{GGA}, u} \boldsymbol{\nabla}\rho(\mathbf{r}) ~
    \chi_{\mu} \chi_{\nu} \left( \mathbf{B} \crossproduct  \mathbf{R}_{\mu \nu} \right)^{B_p} d\mathbf{r} \\
    & + \frac{\text{i}}{4c} \int  v^{\text{XC}}_{\text{mGGA}, u} \left[ (\mathbf{R}_{\mu \nu}\crossproduct \mathbf{r})_p ~ \boldsymbol{\nabla} \chi_{\mu} \cdot \boldsymbol{\nabla} \chi_{\nu} \right. \\
    & + \left. \chi_{\mu} [(\mathbf{r}- \mathbf{R}_{\mu}) \crossproduct \boldsymbol{\nabla} \chi_{\nu}]_p
    - \chi_{\nu} [(\mathbf{r} - \mathbf{R}_{\nu}) \crossproduct \boldsymbol{\nabla} \chi_{\mu}]_p \right] d\mathbf{r}
\end{split}
\label{eq:xc-b}
\end{equation}
where $v^{\text{XC}}_{\text{LSDA}}$ etc.\ denote the exchange-correlation (XC) functional ingredient of the local spin density approximation
(LSDA), generalized gradient approximation (GGA), and the \textit{meta}-GGA rung in the (regularized)
Scalmani--Frisch \cite{Scalmani.Frisch:New.2012, Egidi.Sun.ea:Two-Component.2017,
Komorovsky.Cherry.ea:Four-component.2019, Franzke.Schosser.ea:Efficient.2024, Franzke.Pausch.ea:Application.2025}
or the canonical formalism. \cite{Kubler.Hock.ea:Density.1988, Van-Wullen:Spin.2002,
Armbruster.Weigend.ea:Self-consistent.2008, Desmarais.Komorovsky.ea:spin-orbit.2021}

The occupied-occupied response kernels $\mathbf{b}^{{\text{kernel}}}$ require the response
due to the perturbed overlap matrix as
\begin{equation}
O_{ij}^{B_p} = - \frac{1}{2} S_{ij}^{B_p} \text{.}
\end{equation}
This allows us to compute the Coulomb and exact exchange contributions directly
based on perturbed density matrices of the occupied-occupied tensor space.
For the DFT exchange-correlation kernel, the density variables and their 
GIAO response are formed as outlined in detail in (2c) literature.
\cite{Maximoff.Scuseria:Nuclear.2004,Olejniczak.Bast.ea:simple.2012,Franzke.Holzer:Exact.2023}
We, however, note that the linear coupling of the non-collinear 
kernel is to be used.\cite{Franzke.Holzer:Application.2026}

\subsection{Kramers-Pair Physical-Spin Map}
\label{subsec:kramers-spin-map}

To proceed with the extraction of the required tensors, our components, 
which are obtained from the constructed tangents, need to be 
re-aligned with the physical spin. For an odd-electron determinant 
$|\Phi_+\rangle$, we define its time-reversal symmetric partner 
by $|\Phi_-\rangle=\hat\Theta|\Phi_+\rangle$. Their occupied overlap is
odd-dimensional and skew-symmetric. The transition spin is therefore obtained 
from its cofactor contraction. After phase fixing, define
\begin{equation}
  m_i=\langle\Phi_+|\hat S_i|\Phi_-\rangle,
  \qquad
  \mathsf K_{\rm raw}=
  \begin{pmatrix}
    2\operatorname{Re}\mathbf m &
   -2\operatorname{Im}\mathbf m &
    2\langle\hat{\mathbf S}\rangle
  \end{pmatrix}.
  \label{eq:kramers-raw-spin-map}
\end{equation}
The columns of $\mathsf K_{\rm raw}$ refer to the local pseudospin axes
$(\mathbf e_1,\mathbf e_2,\mathbf n)$. With
$\mathsf F=(\mathbf e_1,\mathbf e_2,\mathbf n)$, 
the full Cartesian map is
\begin{equation}
  \langle\hat{\mathbf S}\rangle
  =\mathsf K_S\widetilde{\mathbf S},
  \qquad
  \mathsf K_S=\mathsf K_{\rm raw}\mathsf F^{\mathrm T},
  \qquad |\widetilde{\mathbf S}|=\frac12.
  \label{eq:kramers-full-spin-map}
\end{equation}
The full map retains transverse mixing, rhombicity, and longitudinal
components. We note that the generally non-integral physical-spin magnitude
$|\langle\hat{\mathbf S}\rangle|$ does not define a spin quantum number.

\subsection{Extraction of the EPR Hyperfine and g Tensors}
\label{sec:extraction}

Of the three relaxed tangents only the transverse pair moves the spin according to
Eq.~\eqref{eq:full-coords}, $\mathbf J_\mu^{\mathrm T}\mathbf t_3=0$ for
$\mu=1,2$. The reference and the two transverse tangent densities therefore
supply the complete set of spin columns,
\begin{equation}
  \mathsf S_{\rm resp}=
  \begin{pmatrix}\mathbf s_0&\mathbf d_1&\mathbf d_2\end{pmatrix},
  \qquad
  \mathbf d_a=\operatorname{Tr}
  (\dot{\mathbf D}_a\hat{\mathbf S}).
  \label{eq:spin-coordinate-matrix}
\end{equation}
If non-singular, $\mathsf S_{\rm resp}$ determines general $3\times3$
physical-response tensors $\mathbf g_{\rm phys}$ and
$\mathbf A_{N,\rm phys}$ without axial symmetry. The third tangent contributes
$\mathbf d_3=\mathbf0$ and no additional spin column. 
Again, this matrix defines a local response basis, 
and must not be confused with a spin quantum number.

The linear model $\mathbf m=\mathbf g_{\rm phys}\mathbf s_{\rm phys}$
predicts no magnetic-moment change along $q_3$, although the determinant
changes. A non-zero residual
\begin{equation}
  \Delta_3=\left|\frac{\mathrm d\mathbf m}{\mathrm dq_3}
  -\mathbf g_{\rm phys}\frac{\mathrm d\mathbf s_{\rm phys}}
  {\mathrm dq_3}\right|
  \label{eq:longitudinal-first-order-probe}
\end{equation}
signals first-order failure of the linear spin model. A genuine $C_n$ axis
through $\mathbf n$ forbids $\Delta_3$. Without it, the residual may be non-zero.

The physical-response matrix retains longitudinal amplitude relaxation.
The corresponding effective-pseudospin basis is
\begin{equation}
  \widetilde{\mathsf S}=
  \begin{pmatrix}
    \tfrac12\mathbf n&\mathbf e_1&\mathbf e_2
  \end{pmatrix}.
  \label{eq:kramers-pseudospin-basis}
\end{equation}
The effective-spin basis and physical-response coordinates $q_a$ are distinct.
They are related at constant order by $\mathsf K_S$, without inversion. 

For nucleus $N$, let
${\mathbf h}^{\text{HFC}}_N=\partial\mathbf {H}_{\mathrm{X2C}}/
\partial\mathbf I_N$ be the picture-changed 2c hyperfine matrix, including
all scalar and spin-dependent terms. \cite{Franzke.Yu:Hyperfine.2022}
Its reference expectation value and tangent derivatives form
\begin{align}
  \mathsf C_N^A= & \,
  \begin{pmatrix}
    \mathbf a_{N,0}&\mathbf a_{N,1}&\mathbf a_{N,2}
  \end{pmatrix}, \\
  \mathbf a_{N,0}=& \, \operatorname{Tr}(\mathbf D_0 {\mathbf h}^{\text{HFC}}_N), \\
  \mathbf a_{N,a}=& \, \operatorname{Tr}(\dot{\mathbf D}_a {\mathbf h^{\text{HFC}}_N}).
  \label{eq:hfc-property-matrix}
\end{align}
In the measured physical-spin coordinates, the local tensor is
\begin{equation}
  \mathbf A_{N,\mathrm{phys}}
  =\mathsf C_N^A\mathsf S_{\rm resp}^{-1}.
  \label{eq:a-from-tangents}
\end{equation}
The effective spin interaction is
\begin{equation}
  \mathbf{ H}_{\mathrm{HFC}}
  =\mathbf I_N^{\mathrm T}\mathbf A_{N,\mathrm{eff}}
   \widetilde{\mathbf S},
  \qquad
  \mathbf A_{N,\mathrm{eff}}
  =\mathbf A_{N,\mathrm{phys}}\mathsf K_S.
  \label{eq:kramers-effective-a}
\end{equation}
The X2C nuclear and electronic constants are applied to
$\mathsf C_N^A$ before this solve when $\mathbf A_N$ is expressed in MHz.

For the electronic Zeeman interaction
$\mathbf{H}_{\mathrm{eZ}}=\mu_{\text{B}}\mathbf B^{\mathrm T}
\mathbf g\hat{\mathbf S}$, define the dimensionless magnetic moment
\begin{equation}
  m_p(\mathbf D)=\frac{1}{\mu_{\text{B}}}\frac{\partial E}{\partial B_p}.
\end{equation}
The stationary value and its derivatives along the two transverse relaxed
tangents give
\begin{align}
  \mathsf C^g&=
  \begin{pmatrix}\mathbf m_0&\mathbf c_1&\mathbf c_2\end{pmatrix} \\
  m_{0,p}&=m_p(\mathbf D_0),
  \label{eq:g-property-matrix}\\
  c_{a,p}&=
  \frac{2}{\mu_{\text{B}}}\operatorname{Re}\sum_{ai}
  b^{B_p}_{ai}\kappa^{(a)}_{ai}
  =\left.\frac{\mathrm d m_p[\mathbf D(q_a)]}
  {\mathrm dq_a}\right|_{0}.
  \label{eq:g-tangent-contraction}
\end{align}
The magnetic orbital gradient $\mathbf b^{B_p}$ is the complete 2c GIAO
gradient. So, it contains the X2C \cite{Franzke.Yu:Quasi-Relativistic.2022, Franzke.Holzer:Application.2026}
and overlap terms together with the Coulomb, exact exchange, 
and exchange--correlation kernel contributions. 
On the measured physical-response coordinates, the
one-reference tensor is
\begin{equation}
  \mathbf g_{\mathrm{phys}}
  =\mathsf C^g\mathsf S_{\rm resp}^{-1}.
  \label{eq:g-from-tangents}
\end{equation}
The full Kramers map then gives the effective tensor
\begin{equation}
  \mathbf{H}_{\mathrm{eZ}}
  =\mu_{\text{B}}\mathbf B^{\mathrm T}\mathbf g_{\mathrm{eff}}
   \widetilde{\mathbf S},
  \qquad
  \mathbf g_{\mathrm{eff}}
  =\mathbf g_{\mathrm{phys}}\mathsf K_S.
  \label{eq:kramers-effective-g}
\end{equation}
The effective tensors enter the shielding. The physical tensors are applied for
the validation and curvature subtraction.
Both property matrices use the same normalized tangents, ensuring covariance
under a common non-singular change of transverse coordinates.

If second-order information is retained, the orientation-manifold expansions
\begin{align}
  \mathbf m(\mathbf q)= & \, \mathbf m_0+
  \mathbf c_aq_a+\tfrac12\mathbf M_{ab}q_aq_b, \\
  \mathbf s_{\rm phys}(\mathbf q)= & \, \mathbf s_0+
  \mathbf d_aq_a+\tfrac12\mathbf L^{\rm phys}_{ab}q_aq_b
  \label{eq:tangent-curvatures}
\end{align}
use the same production coordinates as the stiffness and include all
six symmetric pairs $ab\in\{11,12,22,13,23,33\}$.
The relaxed magnetic curvature is
\begin{align}
  \mathbf M_{ab}= & \,\mathbf M_{ab}^{\mathrm{rigid}}+
  \left\langle\mathbf b^{\mathbf B},\mathbf z_{ab}\right\rangle, \\
  \mathbf H\mathbf z_{ab}= & \,-\boldsymbol{\mathcal Q}
  (\mathbf t_a^{\mathrm{orb}},\mathbf t_b^{\mathrm{orb}}), \\
  \mathbf J^{\mathrm T}\mathbf z_{ab}= & \, \mathbf 0,
  \label{eq:relaxed-curvature}
\end{align}
where $\boldsymbol{\mathcal Q}$ is the second derivative of the orbital
stationarity condition. The constraint uses the full $\mathbf J$ of
Eq.~\eqref{eq:full-coords}.
Mixed blocks follow by polarization,
\begin{equation}
  \mathbf z_{ab}=\tfrac12\left(
  \mathbf z[\mathbf t_a+\mathbf t_b]-\mathbf z_{aa}-\mathbf z_{bb}\right),
  \label{eq:curvature-polarization}
\end{equation}
so six sampled directions and constrained solves determine all six pairs.
Longitudinal relaxation can reach several tens of percent of the rigid term
and must also be included. The same relaxed density path gives the spin curvature,
\begin{equation}
  \mathbf L^{\rm phys}_{ab}=\operatorname{Tr}
  (\mathbf D_{,ab}\hat{\mathbf S}).
  \label{eq:physical-spin-curvature}
\end{equation}
The zeroth-order and first-order equations determine
$\mathbf g_{\rm phys}$ uniquely. The second-order information is retained as
the irreducible physical-response remainder
\begin{equation}
  \mathbf R^g_{ab}=\mathbf M_{ab}
  -\mathbf g_{\mathrm{phys}}\mathbf L^{\mathrm{phys}}_{ab},
  \quad
  \mathbf m(\mathbf q)=
  \mathbf g_{\mathrm{phys}}\mathbf s_{\mathrm{phys}}(\mathbf q)
  +\frac12\mathbf R^g_{ab}q_aq_b.
  \label{eq:full-k-g-rank2-completion}
\end{equation}
Using $\mathbf g_{\mathrm{eff}}$ in this subtraction would mix the effective-pseudospin
and physical-response coordinates and is incorrect.

For the hyperfine property, the corresponding curvature is obtained directly
from the complete rigid-plus-relaxed density curvature,
\begin{equation}
  \mathbf C^A_{N,ab}=\operatorname{Tr}
  \left(\mathbf D_{,ab}{\mathbf h}_N^{\text{HFC}}\right),
  \qquad
  \mathbf D_{,ab}=\mathbf D^{\rm rigid}_{,ab}
                 +\mathbf D_{,\kappa}\mathbf z_{ab}.
\end{equation}
The physical-response quadratic completion is
\begin{align}
  \mathbf C^A_N(\mathbf q)={}&
  \mathbf A_{N,\mathrm{phys}}\mathbf s_{\mathrm{phys}}(\mathbf q)
  +\frac12\mathbf R^A_{N,ab}q_aq_b,\\
  \mathbf R^A_{N,ab}={}&\mathbf C^A_{N,ab}
  -\mathbf A_{N,\mathrm{phys}}\mathbf L^{\mathrm{phys}}_{ab}.
  \label{eq:full-k-hfc-rank2-completion}
\end{align}

\subsection{Assembly of the Paramagnetic Shielding}
\label{sec:assembly}

The response within the effective Kramers manifold is
\begin{equation}
  \boldsymbol\sigma_N^{\mathrm{para}}(T)
  =-\frac{\mu_{\text{B}}}{\gamma_N\hbar}\,
  \mathbf g_{\mathrm{eff}}\,\boldsymbol\chi_{\widetilde S}(T)\,
  \mathbf A_{N,\mathrm{eff}}^{\mathrm T}.
  \label{eq:pnmr-general}
\end{equation}
Here $\mathbf A_{N,\mathrm{eff}}$ is expressed in energy units, $\gamma_N$ is
the nuclear gyromagnetic ratio, $\mu_{\text{B}}$ denotes Bohr's magneton, and
$\boldsymbol\chi_{\widetilde S}$ is the dimensionless-spin
susceptibility in inverse-energy units; $\hbar$ denotes the reduced
Planck constant. For the unsplit doublet,
\begin{equation}
  \boldsymbol\chi_{\widetilde S}(T)
  =\frac{1}{4k_{\text{B}}T}\mathbf1,
  \qquad
  \boldsymbol\sigma_N^{\mathrm{para}}(T)
  =-\frac{\mu_{\text{B}}}{\gamma_N\hbar}
    \frac{\mathbf g_{\mathrm{eff}}
          \mathbf A_{N,\mathrm{eff}}^{\mathrm T}}{4k_{\text{B}}T},
  \label{eq:pnmr-curie}
\end{equation}
where $k_{\text{B}}$ is Boltzmann's constant.

The local orientation energy is
\begin{equation}
  E(\mathbf q)=E_0+
  \frac12\mathbf q^{\mathrm T}\mathbf K\mathbf q .
  \label{eq:orientation-energy-bowl}
\end{equation}
For a symmetric orientational distribution,
$\langle q_a\rangle_T=0$ and
$\mathcal C_{ab}(T)=\langle q_aq_b\rangle_T$, giving
\begin{equation}
  \begin{aligned}
  \langle\mathbf m\rangle_T
    &=\mathbf m_0+\frac12\mathbf M_{ab}\mathcal C_{ab}(T)+\cdots,\\
  \langle\mathbf C_N^A\rangle_T
    &=\mathbf C_{N,0}^A+
      \frac12\mathbf C^A_{N,ab}\mathcal C_{ab}(T)+\cdots .
  \end{aligned}
  \label{eq:thermal-curvature-average}
\end{equation}
A classical harmonic distribution gives
$\boldsymbol{\mathcal C}(T)=k_{\text{B}}T\mathbf K^{-1}$, but the stiffness alone
determines neither excitation gaps nor populations.

The missing dynamics is supplied by the antisymmetric Berry
metric $\boldsymbol\Omega$ of the relaxed tangents. To quadratic order, their
collective Lagrangian is
\begin{equation}
  L=\frac{\hbar}{2}\mathbf q^{\mathrm T}
  \boldsymbol\Omega\dot{\mathbf q}
  -\frac12\mathbf q^{\mathrm T}\mathbf K\mathbf q,
  \qquad
  \Omega_{ab}=2\,\operatorname{Im}
  \langle\partial_a\Phi|\partial_b\Phi\rangle .
  \label{eq:collective-lagrangian}
\end{equation}
The convention for $\boldsymbol\Omega$ must use the same normalized
orientation coordinates as $\mathbf K$. For a harmonic solution
$\mathbf q(t)=\mathbf v e^{\mathrm i\omega t}$,
\begin{equation}
  (\mathbf K-\mathrm i\epsilon\boldsymbol\Omega)\mathbf v=0,
  \qquad \epsilon=\hbar\omega .
  \label{eq:collective-mode-equation}
\end{equation}
In these coordinates $\boldsymbol\Omega$ is $3\times3$ antisymmetric and is
again the Watanabe--Brauner form of
Eq.~\eqref{eq:analytic-orientation-jacobian}. That is, it is of rank two
with kernel $\mathbf n$, transverse block
$\Omega_{12}\left(\begin{smallmatrix}0&1\\-1&0
\end{smallmatrix}\right)$, third row and column zero. Thus, $q_3$ has no
conjugate partner and supports no propagating mode. Restricted to the
transverse pair,
\begin{equation}
  \epsilon=\frac{\sqrt{\det\mathbf K_\perp}}{|\Omega_{12}|}.
  \label{eq:collective-mode-energy}
\end{equation}
The transverse pair describes one positive-frequency precessional mode.
The unpaired third coordinate contributes to the static energy and property
curvatures but has no independent dynamics. Harmonic quantization requires a
stable SOC-lifted minimum with positive-definite $\mathbf K$ and non-zero
$\Omega_{12}$.

Quantization of this one mode gives the covariance
\begin{equation}
  \boldsymbol{\mathcal C}_\perp(T)=
  \frac{\epsilon}{2}\coth\!\left(\frac{\epsilon}{2k_{\text{B}}T}\right)
  \mathbf K_\perp^{-1},
  \label{eq:quantum-mode-covariance}
\end{equation}
where $\perp$ denotes the transverse block, the only one quantized.
With the transverse coordinates fixed, an optional classical model for
the third coordinate is
$\mathcal C_{33}(T)=k_{\text{B}}T/K_{33}$, which vanishes with $T$ instead of saturating
at a zero-point value.
As the KS reference represents the minimum rather than a separately
quantized zero-point average, we retain it exactly and use the thermal excess
\begin{align}
  \Delta\boldsymbol{\mathcal C}_\perp(T)
  = & \boldsymbol{\mathcal C}_\perp(T)-\boldsymbol{\mathcal C}_\perp(0)
  =\epsilon n_B(\epsilon,T)\mathbf K_\perp^{-1},  \label{eq:thermal-excess-covariance} \\
  n_B= & \frac{1}{e^{\epsilon/(k_{\text{B}}T)}-1}.
\end{align}
This vanishes as $T\to0$ and approaches $k_{\text{B}}T\mathbf K_\perp^{-1}$ at high
temperature. The $13$, $23$, and $33$ remainders remain part of the reported
response but contribute to thermal averages only if a corresponding
$\Delta\mathcal C_{ab}$ is supplied, which is connected to the Berry mode
of equation~\ref{eq:collective-mode-energy}.

To exclude curvature already represented by the constant physical-response
tensors, contract the remainders of
Eqs.~\eqref{eq:full-k-g-rank2-completion} and
\eqref{eq:full-k-hfc-rank2-completion} with the thermal excess covariance,
\begin{equation}
  \Delta\mathbf r^g(T)=\frac12\mathbf R^g_{ab}
  \Delta\mathcal C_{ab}(T),
  \qquad
  \Delta\mathbf r_N^A(T)=\frac12\mathbf R^A_{N,ab}
  \Delta\mathcal C_{ab}(T).
  \label{eq:mode-property-average}
\end{equation}
These collective mode corrections vanish when 
constant physical-response tensors fully describe both properties.
In the effective-pseudospin basis, these vector increments are represented by
minimal longitudinal tensor updates along
$\widetilde{\mathbf s}_0=\mathbf n/2$,
\begin{equation}
  \Delta\mathbf g_{\mathrm{eff}}(T)=
  \frac{\Delta\mathbf r^g(T)\widetilde{\mathbf s}_0^{\mathrm T}}
       {|\widetilde{\mathbf s}_0|^2},
  \qquad
  \Delta\mathbf A_{N,\mathrm{eff}}(T)=
  \frac{\Delta\mathbf r_N^A(T)\widetilde{\mathbf s}_0^{\mathrm T}}
       {|\widetilde{\mathbf s}_0|^2}.
  \label{eq:collective-curvature-map}
\end{equation}
This mapping fixes the action along $\widetilde{\mathbf s}_0$ while leaving the transverse
columns of the reference tensors unchanged. Through first order in the
curvatures, the resulting shielding is
\begin{equation}
  \begin{aligned}
  \boldsymbol\sigma_N^{\mathrm{para}}(T)
  =-\frac{\mu_{\text{B}}}{\gamma_N\hbar}\big[&
  \mathbf g_{\mathrm{eff}}\boldsymbol\chi_{\widetilde S}(T)
     \mathbf A_{N,\mathrm{eff}}^{\mathrm T}
  +\Delta\mathbf g_{\mathrm{eff}}(T)\boldsymbol\chi_{\widetilde S}(T)
     \mathbf A_{N,\mathrm{eff}}^{\mathrm T}\\
  &+\mathbf g_{\mathrm{eff}}\boldsymbol\chi_{\widetilde S}(T)
     \Delta\mathbf A_{N,\mathrm{eff}}^{\mathrm T}(T)
  \big]+\mathcal O(\text{curvature}^2) .
  \end{aligned}
  \label{eq:pnmr-collective-mode}
\end{equation}
The product of both tensor increments is of second order in the curvature
and is omitted. The temperature enters through $n_B$ and
$\boldsymbol\chi_{\widetilde S}(T)$. The curvatures and mode energy are fixed.
To avoid double counting, the mode correction must exclude levels already
included in the effective-spin Hamiltonian.
If $\mathbf A_{N,\mathrm{eff}}/h$ is expressed as a frequency in Hz, 
the prefactor in
Eq.~\eqref{eq:pnmr-general} becomes $h\mu_{\text{B}}/(g_N\mu_N)$ including
the nuclear g-factor $g_N$ and Planck's constant $h$. MHz must first be converted to Hz, and the final
dimensionless shielding is multiplied by $10^6$ for the ppm scale.

\subsection{Assembly of Magnetizabilities}
\label{sec:magnetizability}
The magnetizability is assembled from the same ingredients as the pNMR shielding tensor. 
Its temperature-independent part is the orbital magnetizability tensor defined as
$\boldsymbol\xi=-\partial^2E/\partial\mathbf B\,\partial\mathbf B$ of the
orientation-constrained reference, i.e.\ with the Goldstone modes projected out, so that
the spin response is not counted twice. The Curie-type spin part follows from the
effective g-tensor and the pseudospin susceptibility of
Eq.~\eqref{eq:pnmr-general},
\begin{equation}
  \boldsymbol\chi(T)=\boldsymbol\xi
  +\mu_{\text{B}}^2\,\mathbf g_{\mathrm{eff}}\,
  \boldsymbol\chi_{\widetilde S}(T)\,\mathbf g_{\mathrm{eff}}^{\mathrm T},
  \label{eq:magnetizability}
\end{equation}
with the molar value $\chi_M=N_{\text{A}}\mu_0\,\mathrm{tr}(\boldsymbol\chi)/3$,
where $N_{\text{A}}$ denotes Avogadro's constant and $\mu_0$ the vacuum magnetic permeability.
The collective-mode correction enters through the same replacement as for the shielding,
$\mathbf g_{\mathrm{eff}}\to\mathbf g_{\mathrm{eff}}+\Delta\mathbf g_{\mathrm{eff}}(T)$
of Eq.~\eqref{eq:collective-curvature-map}, and the product second order in the curvature 
is again omitted.

Similar to the orbital contribution of the shielding tensor in Eq.~\eqref{eq:orbital-shielding-definition},
the orbital contribution $\boldsymbol\xi$ requires the second-order integral derivatives
with respect of the magnetic field to be contracted with the ground-state density matrix
and the first-order integral derivatives to be contracted with the perturbed density
matrix of the magnetic field. Here, the derivatives arise due to the magnetic perturbation
in the operators and the GIAOs.

\subsection{Implementation}
\label{subsec:implementation}

The two-component methodology outlined herein was implemented into TURBOMOLE.
\cite{Franzke.Holzer.ea:TURBOMOLE.2023, Franzke.Sharma.ea:Recent.2026, TURBOMOLE}
We apply X2C with the superposition of atomic potentials (SAP) to account for the two-electron
picture-change correction. \cite{Surjuse.Valeev:SAP-X2C.2026, Franzke.Holzer:Application.2026}
The SAP-X2C Hamiltonian and all its derivatives except for the ones for the
magnetizability can be used from our previous work. \cite{Franzke.Holzer:Application.2026,
Franzke.Holzer:Exact.2023, Franzke.Yu:Hyperfine.2022, Franzke.Yu:Quasi-Relativistic.2022}
Magnetizabilities require the second-order derivative of the Hamiltonian
and can be implemented straightforwardly based on the existing second-order
machinery. \cite{Franzke.Mack.ea:NMR.2021, Franzke.Holzer:Exact.2023, InB-Franzke:2025}
The non-Hermitian Davidson algorithm to solve the general open-shell response
equations was described previously.\cite{Holzer.Klopper:Ionized.2019,Kehry.Franzke.ea:Quasirelativistic.2020}
In case of stalling CPKS convergence, a minimum residual sub-space solver is used
as a fallback.
Moreover, the conductor-like screening model \cite{Schafer.Klamt.ea:COSMO.2000,
Pausch:Consistent.2024} (COSMO) is directly available throughout, as it is of
the same symmetry structure as the Coulomb integrals. That is, it requires
both the GIAO-based integrals for the RHS of the response equations
\cite{Reiter.Mack.ea:Calculation.2018} and the response terms for the LHS.

The response part proceeds in two stages with the same
2c Hessian $\mathbf H$. First, three CPKS equations $\mathbf H\mathbf Y=\mathbf J$ 
are solved, one for each orientation covector $\mathbf J_a$ of 
Eqs.~\eqref{eq:analytic-covectors} and \eqref{eq:third-covector}.
They yield the compliance $\mathbf R=\mathbf J^{\mathrm T}\mathbf Y$, 
the stiffness $\mathbf K=\mathbf R^{-1}$, and
the relaxed tangents $\mathbf T=\mathbf Y\mathbf R^{-1}$ 
(Eq.~\eqref{eq:relaxed-tangents}), from which
the tangent densities, the Berry metric, as well as the g and 
hyperfine tensors follow by contraction
(\eqref{eq:a-from-tangents} and Eqs.~\eqref{eq:g-from-tangents}) 
without any further response calculation.
Second, nine further CPKS equations are solved with the 
same $\mathbf H$. Three for the components of the magnetic field,
$\mathbf H\mathbf x^{B_p}=-\mathbf b^{B_p}$, 
which are brought to the constrained solution of the orbital shielding
and magnetizability by the Schur correction of 
Eq.~\eqref{eq:schur-projection}. Six constrained solves are required for the second-order 
orientation responses $\mathbf z_{ab}$ of Eq.~\eqref{eq:relaxed-curvature}
(the three diagonal directions and three sums of 
tangents, with the mixed blocks obtained by the polarization of
Eq.~\eqref{eq:curvature-polarization}). The latter six are only needed for 
the curvature corrections and the collective-mode correction.

The convergence of the Kohn--Sham response equations and the validity 
of our framework requires a well
converged ground-state SCF solution, i.e.\ not only a stationary point 
but a minimum is strictly
required. Therefore, we generalized the augmented Roothaan-Hall (ARH) solver
to the two-component formalism with fixed integer occupations, fixed
fractional occupations as well varying fractional occupations for
smearing and annealing procedures to find the proper stationary
ground state solution. The 2c ARH theory is described in the appendix
of the present work. There, we also describe the extension of popular
population analyses from the 1c to the 2c framework to aid in the
interpretation of results.

\section{Computational Methods}
\label{sec:comput}

All calculations were performed with a development version of \textsc{Turbomole}.
\cite{Franzke.Holzer.ea:TURBOMOLE.2023, Franzke.Sharma.ea:Recent.2026, TURBOMOLE}
Relativistic effects were treated with the spin--orbit two-component exact two-component 
Hamiltonian in an all-electron formulation, using the diagonal local approximation
to the unitary decoupling transformation (DLU-X2C), \cite{Peng.Middendorf.ea:efficient.2013, 
Franzke.Middendorf.ea:Efficient.2018} a Gaussian finite-nucleus model for both the
scalar and the vector potential, \cite{Visscher.Dyall:Dirac-Fock.1997, Franzke.Weigend:NMR.2019}
and a superposition of atomic potentials (SAP) to account for the two-electron contributions
to the X2C Hamiltonian, \cite{Surjuse.Valeev:SAP-X2C.2026, Franzke.Holzer:Application.2026}
with the SAP basis of Ref.~\citenum{Lehtola.Visscher.ea:Efficient.2020}.
A Kramers-unrestricted non-collinear 2c Kohn--Sham framework with the PBE0 hybrid
functional~\cite{Perdew.Burke.ea:Generalized.1996, Adamo.Barone:Toward.1999}
was employed. Exact exchange was evaluated seminumerically, \cite{Holzer:improved.2020}
while MARI-J was used for the Coulomb part. \cite{Sierka.Hogekamp.ea:Fast.2003,
Armbruster.Weigend.ea:Self-consistent.2008}
We stress again that Eq.~\ref{eq:2c-density} is assumed to be
the ground state density. A failure to use the true ground state density 
will lead to gauge-variance, as further correction terms would occur
in Sec.~\ref{sec:2c-nonkramers-nmr}. Given that the symmetry breaking
leads to very small Hessian eigenmodes, we performed a stability 
analysis for every KS reference to verify its true ground state nature.
The 2c ARH solver outlined in the appendix was able to yield
the correct state in most cases, while we specifically warn that 
direct inversion in iterative subspace (DIIS) led to an excited state, 
usually $10^{-4}$ to $10^{-2}$\,$E_{\text{h}}$ above the ground state, in 
\textit{every} actinide-containing molecule, independent of the starting point. 
The g and hyperfine tensors, the magnetizabilities, the shieldings, and the 
spin--orbit orientation curvature (Berry metric and stiffness) were obtained with 
\texttt{mpshift} on the converged reference, with the Kramers chart
(\texttt{\$spinrotgderiv nosym kramers}) for all systems.

For Ru, Cl, and S the all-electron x2c-TZVPall-2c basis was used, 
\cite{Pollak.Weigend:Segmented.2017}
and for H, C, N, O the x2c-TZVPall-2c-s basis was used.
\cite{Franzke.Tress.ea:Error-consistent.2019}
Here, tailored MARI-J/RI-J basis sets are available.
\cite{Pollak.Weigend:Segmented.2017, Franzke.Spiske.ea:Segmented.2020}
For U and Np, the all-electron Dyall-VTZ basis was employed in fully uncontracted form.
\cite{Dyall:Relativistic.2007A, Dyall-Repo} Auxiliary basis sets for
the Dyall-VTZ orbital basis set were generated automatically~\cite{Lehtola:Automatic.2023, Franzke.Sharma.ea:Recent.2026} and are given in the Supporting Information.
The solvent was modeled with COSMO~\cite{Schafer.Klamt.ea:COSMO.2000, Pausch:Consistent.2024}
for the Ru complexes in dimethylformamide (DMF, $\varepsilon_{\text{r}}=36.71$)
and water ($\varepsilon_{\text{r}}=80.1$). Furthermore, we performed calculations
for the gas phase. For the actinide complexes, COSMO with the settings for water
was used throughout to allow for a comparison with the experiment.
A grid of size 4a was used. \cite{Treutler.Ahlrichs:Efficient.1995, Franzke.Tress.ea:Error-consistent.2019}
The SCF procedure was converged to an energy change below $10^{-7}\,E_{\text{h}}$ and
an root mean square (RMS) density change below $10^{-6}$. The molecular structures were
optimized at the same level with convergence thresholds of $10^{-6}\,E_{\text{h}}$ for
the energy and $10^{-3}\,E_{\text{h}}/a_0$ for the
maximum Cartesian gradient. \cite{Franzke.Sharma.ea:Recent.2026}


\begin{table*}[t]
  \centering
  \caption{Principal $g$ values of $[\mathrm{RuCl_4(DMSO)(L)}]^-$ from this work (2c SAP-X2C-PBE0), 
  calculated literature values \cite{Novotny.Sojka.ea:Interpreting.2016, Chyba.Hruzikova.ea:Nature.2023}
  and experiment. \cite{Webb.Walsby:Albumin.2015, Novotny.Sojka.ea:Interpreting.2016, paula2000epr}
  The components and g-shift are defined as $g_\perp=(g_1+g_2)/2$, $g_\parallel=g_3$, and $\Delta g=g_\perp-g_\parallel$.
  PBS indicates the phosphate-buffered saline solution.}
  \label{tab:g-nami}
  \footnotesize
  \setlength{\tabcolsep}{10pt}
  \begin{tabular}{l l c c c c c c}
    \hline\hline
    L & Source & $g_1$ & $g_2$ & $g_3$ & $g_\perp$ & $g_\parallel$ & $\Delta g$ \\
    \hline
    {}[RuCl$_4$(DMSO)(Im)]$^-$ (NAMI-A) & this work, gas & 2.303 & 2.297 & 1.882 & 2.300 & 1.882 & 0.418 \\
     & this work, COSMO (DMF) & 2.388 & 2.345 & 1.753 & 2.367 & 1.753 & 0.614 \\
     & this work, COSMO (water) & 2.392 & 2.346 & 1.747 & 2.369 & 1.747 & 0.622 \\
    \hline
    {}[RuCl$_4$(DMSO)(py)]$^-$ & this work, gas & 2.318 & 2.246 & 1.879 & 2.282 & 1.879 & 0.403 \\
     & this work, COSMO (DMF) & 2.379 & 2.306 & 1.793 & 2.342 & 1.793 & 0.550 \\
     & this work, COSMO (water) & 2.380 & 2.309 & 1.789 & 2.344 & 1.789 & 0.555 \\
    \hline
    4-Me-py & SO-ZORA \cite{Novotny.Sojka.ea:Interpreting.2016} & 2.434 & 2.399 & 1.924 & 2.417 & 1.924 & 0.493 \\
    4-CN-py & SO-ZORA \cite{Novotny.Sojka.ea:Interpreting.2016} & 2.419 & 2.374 & 1.934 & 2.397 & 1.934 & 0.463 \\
    4-Me-py & SO-ZORA (3a) \cite{Chyba.Hruzikova.ea:Nature.2023} & 2.370 & 2.340 & 1.850 & 2.355 & 1.850 & 0.505 \\
    \hline
    NAMI-A (Im) & exp., PBS, 20\,K\cite{Webb.Walsby:Albumin.2015} & 2.47 & 2.47 & 1.72 & 2.47 & 1.72 & 0.75 \\
    NAMI-A-H$_2$O & exp., PBS, 20\,K\cite{Webb.Walsby:Albumin.2015} & 2.30 & 2.30 & 1.88 & 2.30 & 1.88 & 0.42 \\
    {}[RuCl$_4$(DMSO)(py)]$^-$ & exp., powder, $-160\,^\circ$C\cite{paula2000epr} & {\text{--}} & {\text{--}} & {\text{--}} & 2.42 & 1.81 & 0.61 \\
    {}[RuCl$_4$(DMSO)$_2$]$^-$ & exp., PBS, 20\,K\cite{Novotny.Sojka.ea:Interpreting.2016} & 2.43 & 2.43 & 1.77 & 2.43 & 1.77 & 0.66 \\
    \hline\hline
  \end{tabular}
\end{table*}

\section{Results \& Discussion}
\label{sec:results}

\subsection{EPR g-tensors}
\label{sec:results-g}

\subsubsection{NAMI-A and its pyridine analogue}

Table~\ref{tab:g-nami} collects the newly calculated principal $g$ values for the gas phase
and for COSMO(DMF) and COSMO(water) (each medium with its own geometry optimization),
the spin--orbit zeroth-order regular approximation (SO-ZORA/PBE0) values of
Ref.~\citenum{Novotny.Sojka.ea:Interpreting.2016} for the 4-substituted pyridine analogues,
as well as the experimentally determined EPR parameters.

Going from the gas phase to either solvent raises $g_\perp$ and lowers $g_\parallel$ for both ligands, 
so that the anisotropy grows by 0.15--0.20. DMF and water give the same tensor within 0.006 in the principal values.
We relate this mostly to geometrical changes in the ground state, outlining the importance 
of the underlying structures to the obtained results. In detail, the solvent lengthens the Ru--S bond 
(imidazole: 2.281\,\AA{} in the gas phase, 2.315/2.317\,\AA{} in DMF/water; pyridine: 2.236, 2.260, 2.261\,\AA) 
and the Ru--N bond by 0.01--0.02\,\AA. The gas-phase imidazole tensor is nearly axial ($g_1-g_2=0.007$), 
whereas all other tensors are slightly rhombic (0.04--0.07).

For NAMI-A, the water calculation, which matches the aqueous experiment, is 0.10 low in $g_\perp$, 
0.03 high in $g_\parallel$ and 0.13 low in the anisotropy. The gas-phase tensor deviates significantly. 
Therefore, the implicit solvent changes the tensor towards the experiment, however, DMF and water are indistinguishable 
in our calculations, aligning with the only minor geometrical changes observed for these solvent.
As a rule of thumb, permittivity values between 20 and 100 lead to very similar result for geometries.
The calculated tensors are closer to the measured anisotropy than the previously reported SO-ZORA values 
of the pyridine analogues. Here, the ZORA tensors slightly overestimate $g_\parallel$ 
(1.85--1.93 against 1.81 measured) and underestimate the anisotropy (0.46--0.51 against 0.61), 
whereas ours give $g_\parallel=1.79$ and $\Delta g=0.55$ for pyridine. We note that the 
measured NAMI-A spectrum was fitted as axial with line widths of 105 and 180\,G, 
\cite{Webb.Walsby:Albumin.2015} which does not exclude a small rhombic splitting of the 
size calculated within our method and the previously reported SO-ZORA method.

\subsubsection{$5f^1$ actinide complexes}
Table~\ref{tab:g-actinide} lists the calculated g-tensors of the 
three U and Np $5f^1$ complexes together with the calculated ground-state
doublet values of the literature using multireference wave-function theory such
as complete active space SCF (CASSCF) or restricted active space SCF (RASSCF) theory
using perturbation theory of second-order for dynamic correlation (PT2). Active
spaces are denoted as usual with (A,B) indicating A active electrons in B active
molecular orbitals. Note that the multireference calculations do not use a 2c formalism
but introduce spin--orbit coupling \textit{a posteriori}. In contrast, the reported SO-ZORA
results are from a 2c DFT approach.

\begin{table}[htbp]
  \centering
  \caption{Principal $g$ values of the ground-state Kramers doublet of the $5f^1$ complexes.
  We compare results from this work (SAP-X2C-PBE0) against wave-function calculations. 
  SO-CASSCF/CASPT2/RASPT2 and restricted SO-ZORA/PBE0 are further listed as taken from 
  Ref.~\citenum{Autillo.Islam.ea:Temperature.2021}.
  PT2-SO results are taken from Ref.~\citenum{Gendron.Pritchard.ea:Magnetic.2014}.
  The literature tensors are axial for the carbonates ($g_\perp$ listed twice here
  as $g_2$ and $g_3$).}
  \label{tab:g-actinide}
  \footnotesize
  \setlength{\tabcolsep}{8pt}
  \begin{tabular}{l l c c c}
    \hline\hline
    Complex & Source & $g_1$ & $g_2$ & $g_3$ \\
    \hline
    NpO$_2$(DPA)$_2^{2-}$ & this work & 1.56 & 0.25 & 0.22 \\
     & SO-CASSCF\cite{Autillo.Islam.ea:Temperature.2021}  & 1.80 & 0.30 & 0.30 \\
     & SO-CASPT2\cite{Autillo.Islam.ea:Temperature.2021}  & 1.60 & 0.50 & 0.50 \\
     & SO-RASPT2\cite{Autillo.Islam.ea:Temperature.2021} & 1.00 & 1.00 & 0.90 \\
     & SO-ZORA/PBE0\cite{Autillo.Islam.ea:Temperature.2021}  & 1.90 & 0.40 & 0.30 \\
    \hline
    UO$_2$(CO$_3$)$_3^{5-}$ & this work & 2.00 & 0.05 & 0.03 \\
     & PT2-SO CAS(1,4)\cite{Gendron.Pritchard.ea:Magnetic.2014} & 3.80 & 0.22 & 0.22 \\
     & PT2-SO CAS(7,10)\cite{Gendron.Pritchard.ea:Magnetic.2014} & 2.12 & 0.01 & 0.01 \\
    \hline
    NpO$_2$(CO$_3$)$_3^{4-}$ & this work & 2.09 & 0.00 & 0.00 \\
     & PT2-SO CAS(1,4)\cite{Gendron.Pritchard.ea:Magnetic.2014} & 1.95 & 0.09 & 0.09 \\
     & PT2-SO CAS(7,10)\cite{Gendron.Pritchard.ea:Magnetic.2014} & 2.07 & 0.01 & 0.01 \\
    \hline\hline
  \end{tabular}
\end{table}

For the carbonates UO$_2$(CO$_3$)$_3^{5-}$ and NpO$_2$(CO$_3$)$_3^{4-}$, the tensors are essentially uniaxial, 
as expected for an $\Omega=\tfrac32$ ($\delta$-type) ground doublet in $D_{3\text{h}}$ symmetry.
Thus, the transverse components almost vanish. Our calculated $g_\parallel$ value is
2.00 (U) and 2.09 (Np), against 2.12 and 2.07 from PT2-SO with the CAS(7,10) active space, 
the transverse components ($\le0.05$) are of the  same size as the calculated 0.01--0.02.
The smaller CAS(1,4) space leads to 3.80 for U(V) but 1.95 for Np(VI), which shows how strongly
the value depends on  the mixing of the $\varphi$ and $\delta$ orbitals.
The calculated $g_\parallel\approx2.1$  is well below the free-ion value of a pure
$|\tfrac52,\pm\tfrac32\rangle$ state (2.57) 
and far below that of $|\tfrac52,\pm\tfrac52\rangle$ (4.29),
which Gendron and co-workers explain by the equatorial ligand field and covalency.
\cite{Gendron.Pritchard.ea:Magnetic.2014}

For the dipicolinic acid derivative NpO$_2$(DPA)$_2^{2-}$, our simulated tensor has the same axial pattern 
as SO-ZORA/PBE0, SO-CASSCF, and SO-CASPT2. It is describes by a $\delta$-type ground doublet with the 
large $g$ component along the O-Np-O axis. The transverse components of 0.2--0.3 are due to the lower
equatorial symmetry of the two tridentate ligands. The axial value lies between the CASPT2 and ZORA values. 
For comparison, the RASPT2 calculations with 5$p$ and 5$d$ correlation results in a nearly isotropic tensor, 
i.e.\ a different ground state.

\subsection{Magnetizabilities}
\label{sec:results-chi}

\subsubsection{NAMI-A and its pyridine analogue}
Table~\ref{tab:chi-nami} presents the molar magnetic susceptibility at 298\,K as obtained from the calculated tangents. 
No susceptibility has been measured for these complexes. Therefore, we use the Curie value of 
the effective spin $\tfrac12$ built from the experimental g-tensors as reference values,
$\chi_M=N_{\text{A}}\mu_0\mu_{\text{B}}^2\langle g^2\rangle/(4k_{\text{B}}T)$ with $\langle g^2\rangle=(g_1^2+g_2^2+g_3^2)/3$.
For NAMI-A in PBS, \cite{Webb.Walsby:Albumin.2015} we obtain 2.00, while the pyridine complexes measured 
powder values~\cite{paula2000epr} yield 1.98. Thus, we observe an excellent agreement.

\begin{table}[htbp]
  \centering
  \caption{Molar magnetic susceptibility $\chi_M$ at 298\,K ($10^{-8}$\,m$^3$\,mol$^{-1}$) 
  of $[\mathrm{RuCl_4(DMSO)(L)}]^-$. We list the orbital part $\xi$, the spin part (spin), 
  and the total result ($\chi_M$), as well as the Curie value from the experimental g-tensor and the ratio.
   \cite{Webb.Walsby:Albumin.2015, paula2000epr}}
  \label{tab:chi-nami}
  \footnotesize
  \setlength{\tabcolsep}{6pt}
  \begin{tabular}{l c c c c c}
    \hline\hline
    System & $\xi$ & spin & $\chi_M$ & Curie (exp.\ $g$) & ratio \\
    \hline
    NAMI-A, gas & 0.01 & 1.86 & 1.87 & 2.00 & 0.94 \\
    NAMI-A, DMF & 0.03 & 1.88 & 1.91 & 2.00 & 0.96 \\
    NAMI-A, water & 0.03 & 1.88 & 1.91 & 2.00 & 0.96 \\
    \hline
    py analogue, gas & $-$0.00 & 1.84 & 1.84 & 1.98 & 0.93 \\
    py analogue, DMF & 0.02 & 1.87 & 1.89 & 1.98 & 0.95 \\
    py analogue, water & 0.02 & 1.87 & 1.89 & 1.98 & 0.96 \\
    \hline\hline
  \end{tabular}
\end{table}

As evident from the results, the total value is the pure Curie term of the pseudospin.
The orbital part amounts to only 0.00--0.03. 
It is 4--7\,\% below the Curie value obtained from the measured g-tensors (93--96\,\% of it), 
consistent with the 0.1 deviation of the calculated $g_\perp$.
The gas-phase values are 2--3\,\% below the solvent values, and DMF and water agree. 
The temperature dependence is Curie-like. That is, between 250 and 350\,K, the susceptibility falls from 
2.27 to 1.63 (NAMI-A, water), a ratio of 0.72 that equals $250/350$ for every calculation. So,
$\chi_MT$ is constant.

\subsubsection{NpO$_2$(DPA)$_2^{2-}$: Importance of the collective mode correction}

\begin{figure}[ht]
  \centering
  \includegraphics[width=0.9\columnwidth]{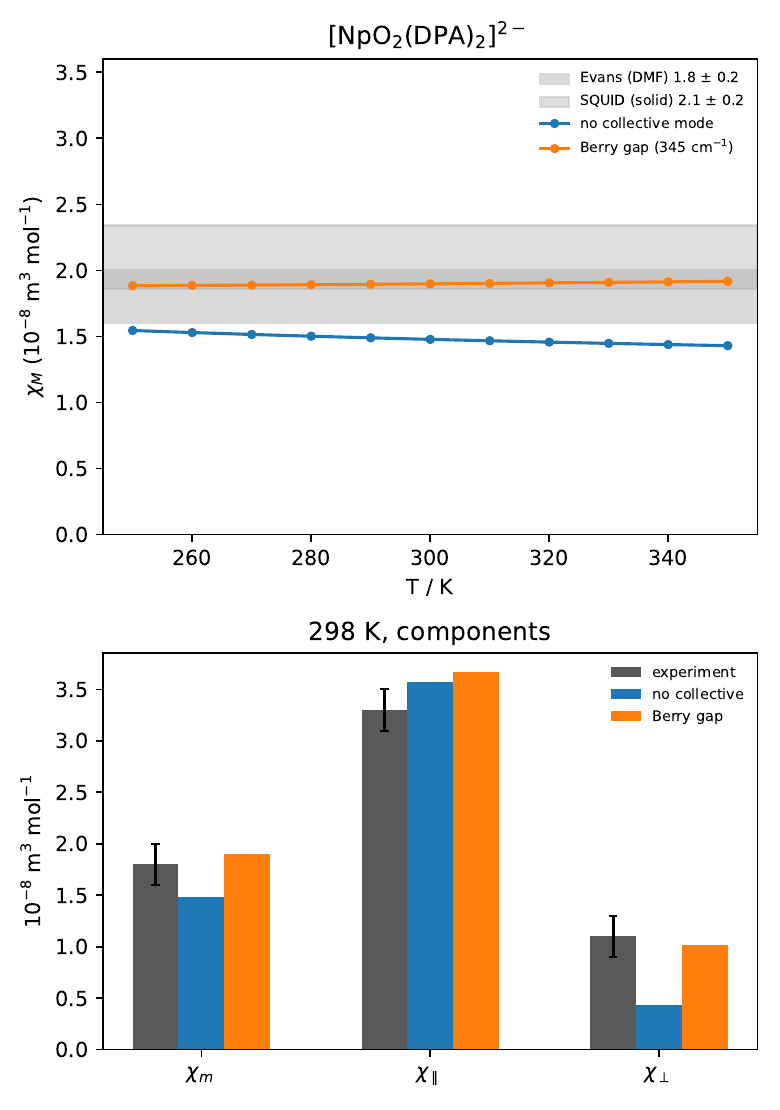}
  \caption{$[\mathrm{NpO_2(DPA)_2}]^{2-}$. Top: temperature dependence of $\chi_M$ (250--350\,K) 
  without and with the Berry-gap collective mode. Here, the gray bands are the values measured at 298\,K 
  (Evans in DMF, SQUID on the solid). \cite{Autillo.Islam.ea:Temperature.2021}
  Bottom: isotropic value as well as axial and equatorial components at 298\,K against the experiment
  of Ref.~\citenum{Autillo.Islam.ea:Temperature.2021}.}
  \label{fig:chi-np}
\end{figure}

For the $5f^1$ complex, the plain assembly is compared with the collective-mode recipes 
in Fig.~\ref{fig:chi-np}. The collective-mode correction is described in Sec.~\ref{sec:assembly}.

The plain assembly leads to $\chi_M(298\,\mathrm K)=1.48$, i.e.\ 82\,\% of the Evans value 
and 70\,\% of the SQUID value. The orbital term $\xi=1.14$ dominates for this case. The axial 
component (3.57, measured 3.3) is within 8\,\% of experiment, however, the equatorial 
component (0.43, measured 1.1) is only 40\,\% of it. Therefore, the anisotropy is overestimated 
($\Delta\chi=3.14$ against 2.2). With the collective mode all three quantities move into 
the experimental range: $\chi_m=1.90$ (Evans $1.8\pm0.2$, SQUID $2.1\pm0.2$), 
$\chi_\perp=1.01$ ($1.1\pm0.2$) and $\chi_\parallel=3.67$ ($3.3\pm0.2$), 
while $\Delta\chi=2.66$ remains too high. The mode
also changes the temperature dependence. Here, the plain susceptibility decreases by 7\,\% 
between 250 and 350\,K (1.54 to 1.43), the collective-mode value is almost constant 
because the growing Curie term is compensated by the thermal correction. Only 298\,K is
available experimentally, revealing that the temperature dependence cannot be tested.
The Berry gap  (345\,cm$^{-1}$) lies between the gaps that reproduce the Evans (430\,cm$^{-1}$)
and the  SQUID value (211\,cm$^{-1}$) and in the range of the first-excited-doublet energies of 
the literature (fits 206--360, \textit{ab initio} calculations 333--820\,cm$^{-1}$). \cite{Autillo.Islam.ea:Temperature.2021}

\subsection{pNMR shifts}
\label{sec:res-shifts}

\subsubsection{Ru(III) NAMI complexes}
\label{sec:res-shifts-nami}

The paramagnetic NMR shifts of the pyridine complex and of the imidazole complex (NAMI-A anion) were calculated 
at 293\,K in the gas phase, DMF, and water. $\delta_{\mathrm{tot}}$ is the shift relative to 
tetreamethylsilane (TMS) in the same medium and $\delta_{\mathrm{para}}$ the paramagnetic shift.

\begin{figure}[b]
  \centering
  \includegraphics[width=0.92\columnwidth]{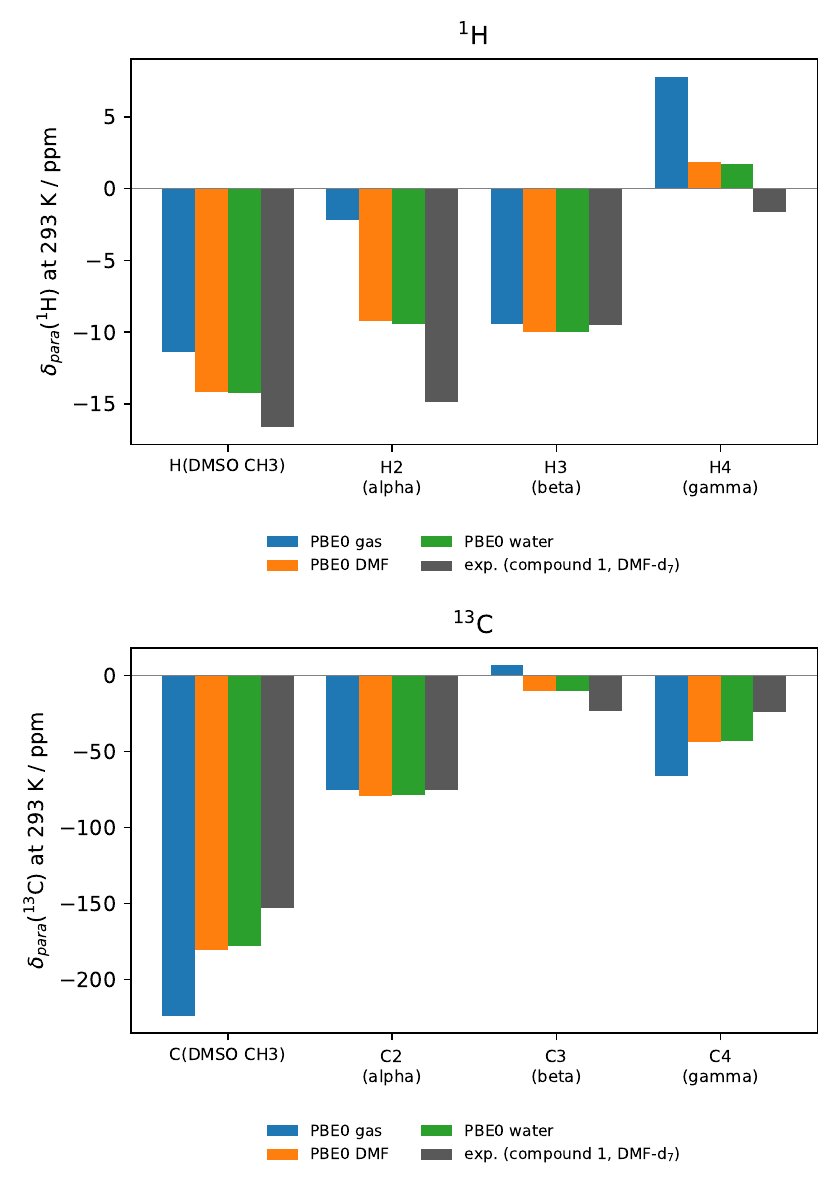}
  \caption{Calculated paramagnetic shifts $\delta_{\mathrm{para}}$ (293\,K) of
    $[\mathrm{RuCl_4(DMSO)(py)}]^-$ in the gas phase, DMF, and water (PBE0)
    compared to the experimental values of compound 1 in DMF-d$_7$.
    \cite{Novotny.Sojka.ea:Interpreting.2016}
    The top panel shows the $^1$H $\delta_{\mathrm{para}}$ shifts, whereas the bottom panel
    illustrates the results for the $^{13}$C $\delta_{\mathrm{para}}$ shifts.}
  \label{fig:nami-py-para}
\end{figure}

In DMF and water, the calculated paramagnetic shifts of the pyridine 
complex are within about 3\,ppm ($^1$H) and 15\,ppm ($^{13}$C) of the 
measured ones as shown in Fig.~\ref{fig:nami-py-para}. The gas-phase shifts are
considerably worse, outlining the again strong dependence on the geometry. 
The best agreement is found for H3 and C2. The {\textalpha}-proton H2 signal is shifted by 5.5\,ppm, 
while for H4 the shift is slightly smaller yet changes the sign. 
The DMSO methyl carbon is shifted by 25--27\,ppm. DMF and water give the same shifts 
within 0.4\,ppm for $^1$H and within 3\,ppm for $^{13}$C.

Turning to the imidazole complex in Table~\ref{tab:nami-shifts-im}, the only experimental shift
is the DMSO $^1$H shift of NAMI-A. Here, the solvent alters the calculated value from $-9.0$ (gas)
to $-11.6$ (DMF) and $-11.7$\,ppm (water), 1.5\,ppm from the measured  $-13.2$\,ppm (gas phase: 4.2\,ppm). 
On the paramagnetic scale the imidazole shifts are of the same 
size and sign as those of pyridine. Here, the carbons next to the coordinated nitrogen (C2, C4) have 
$\delta_{\mathrm{para}}=-119$ and $-65$\,ppm in DMF (pyridine {\textalpha}-carbon: $-79$ calculated, 
$-75$\,ppm measured), and the $^1$H shifts next to the nitrogen are found at $-9.0$ (H2) and $-11.4$\,ppm (H4). 
The orbital part changes much less with the medium than the paramagnetic part. That is, it changes by at most 11\,ppm
for the ring $^{13}$C and 2\,ppm for $^1$H, against up to 46\,ppm and 7\,ppm for $\delta_{\mathrm{para}}$.

\begin{table}[htbp]
  \centering
  \caption{Calculated total and paramagnetic shifts (ppm, 293\,K) of the NAMI-A anion
    $[\mathrm{RuCl_4(DMSO)(Im)}]^-$ in the gas phase, DMF, and water. 
    For the $^1$H, DMSO CH$_3$, a shift of $-$13.2 was found at ambient 
    temperature. No other shifts were reported experimentally at this condition.}
  \label{tab:nami-shifts-im}
  \footnotesize
  \setlength{\tabcolsep}{4pt}
  \begin{tabular}{l ccc ccc}
    \hline\hline
    & \multicolumn{3}{c}{$\delta_{\mathrm{tot}}$} & \multicolumn{3}{c}{$\delta_{\mathrm{para}}$} \\
    \cline{2-4}\cline{5-7}
    Nucleus, Moiety & gas & DMF & water & gas & DMF & water \\
    \hline
    $^1$H, DMSO CH$_3$ & $-$9.0 & $-$11.6 & $-$11.7 & $-$12.1 & $-$14.8 & $-$14.9 \\
    $^1$H, H2 (C2--H) & +3.8 & $-$1.6 & $-$1.7 & $-$4.1 & $-$9.0 & $-$9.2 \\
    $^1$H, H4 (C4--H) & $-$0.3 & $-$4.0 & $-$4.1 & $-$8.1 & $-$11.4 & $-$11.5 \\
    $^1$H, H5 (C5--H) & $-$0.4 & $-$1.8 & $-$1.8 & $-$7.0 & $-$9.2 & $-$9.3 \\
    $^1$H, H1 (N--H) & +2.6 & +1.7 & +1.7 &  $-$5.2 & $-$7.9 & $-$8.0 \\
    $^{13}$C, DMSO CH$_3$ & $-$146 & $-$104 & $-$102 &  $-$192 & $-$148 & $-$147 \\
    $^{13}$C, C2 & +40 & +27 & +27 & $-$105 & $-$119 & $-$119 \\
    $^{13}$C, C4 & +89 & +74 & +74 & $-$54 & $-$65 & $-$65 \\
    $^{13}$C, C5 & +80 & +80 & +80 & $-$32 & $-$43 & $-$44 \\
    \hline\hline
  \end{tabular}
\end{table}

\subsubsection{[AnO$_2$(CO$_3$)$_3$]$^{\,n-}$ (U(V), Np(VI))}
\label{sec:res-shifts-carb}

\begin{figure*}[htbp]
  \centering
  \includegraphics[width=0.86\textwidth]{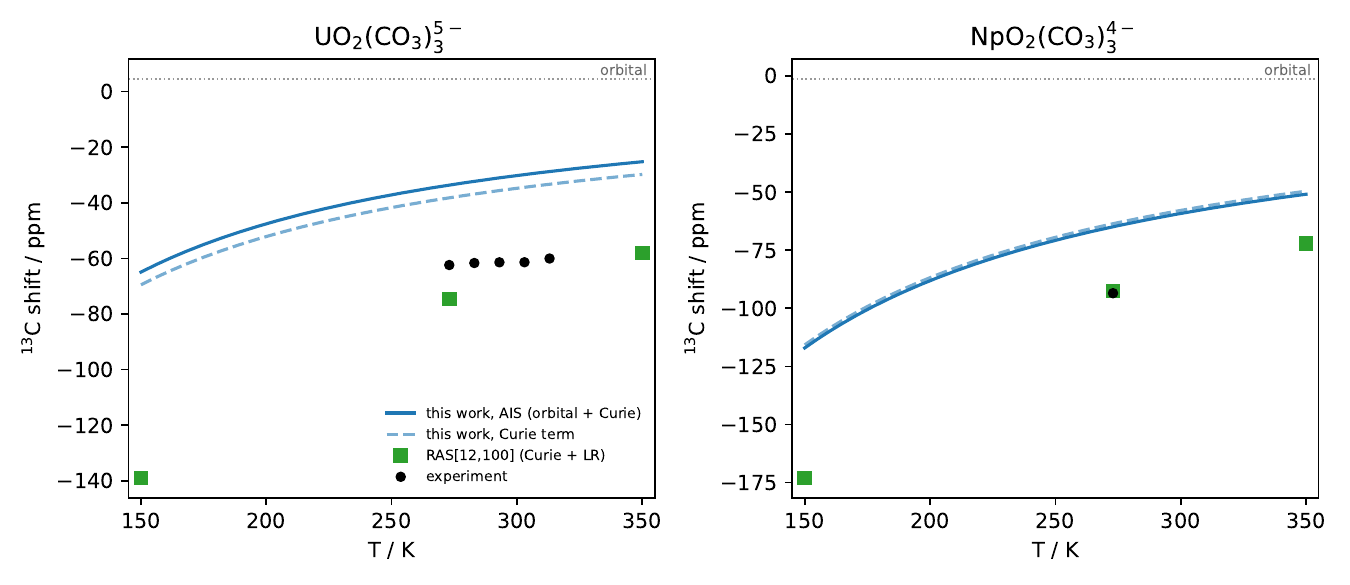}
  \caption{$^{13}$C shift of the carbonates between 150 and 350\,K relative to UO$_2$(CO$_3$)$_3^{4-}$: 
  AIS of SAP-X2C-PBE0 (orbital difference plus Curie term), the Curie term alone, the RAS[12,100] shifts of
  Ref.~\citenum{Gendron.Autschbach:Ligand.2016} (Curie plus LR terms) 
  and experiment (digitized for U(V),\cite{Mizuoka.Grenthe.ea:Structural.2005} 273\,K 
  for Np(VI)\cite{Clark.Hobart.ea:Actinide.1995}). 
  The dotted gray line is the orbital difference to the U(VI) complex UO$_2$(CO$_3$)$_3^{4-}$.}
  \label{fig:carb-T}
\end{figure*}

For the two actinide carbonyl complexes, Fig.~\ref{fig:carb-T} shows the shifts between 150 and 350\,K.
Here, the orbital part is cleanly separated and
only yields small shifts compared to the closed shell UO$_2$(CO$_3$)$_3^{4-}$ reference.

Concerning the temperature dependence, the measured shift of U(V) changes from 
$-62.4$ to $-60.0$\,ppm between 273 and 313\,K, i.e.\ by 4\,\% 
(Table~\ref{tab:shifts-carb-terms}), 
whereas both calculations follow a Curie law and change by 13\,\% (RAS[12,100], 
interpolated: $-74.6$ to $-65$\,ppm, 9.6\,ppm) and 15\,\% (SAP-X2C-PBE0, 4.9\,ppm). 
The temperature dependence of the CAS/RASSCF results is therefore too large compared 
with experiment, by a factor 3.4 relative to the shift and 4 in ppm. 
The calculation of this work shares the Curie-like relative dependence. 
The RAS[12,100] shift of U(V) reaches the measured $-62.4$\,ppm at 
326\,K (338\,K for the 313\,K value) instead of 273\,K, and that of 
Np(VI) at 270\,K. The actinide induced shifts (AIS) of SAP-X2C-PBE0 reaches the measured value at
156\,K for U(V) (161\,K using the 313\,K value) and at 188\,K for Np(VI), 
i.e.\ its Curie constant ($C=\delta_{\mathrm{para}}T=-10424$ and $-17361$\,ppm\,K) 
is too small by a factor 1.75 (U) and 1.45 (Np) at 273\,K. For Np(VI), only 
the 273\,K experiment exists, so the temperature dependence cannot be compared.

\begin{table}[b]
  \centering
  \caption{Terms of the $^{13}$C shift of the carbonates at 273\,K (ppm), relative to UO$_2$(CO$_3$)$_3^{4-}$. We list the
  Curie term, linear response (LR) term describing coupling to the excited states, and total.
  For the wave-function calculations of Ref.~\citenum{Gendron.Autschbach:Ligand.2016} and calculations of this work, we report the
  Curie term $\delta_{\mathrm{para}}$ of the plain assembly, orbital difference of the same calculation, and the actinide induced shifts (AIS).
  The orbital difference is only reported within the present work at the SAP-X2C-PBE0 level. It is, however, very small and amounts to only
  $+4.6$ and $-1.3$\,ppm for UO$_2$(CO$_3$)$_3^{5-}$ and NpO$_2$(CO$_3$)$_3^{4-}$, respectively.}
  \label{tab:shifts-carb-terms}
  \footnotesize
  \setlength{\tabcolsep}{5pt}
  \begin{tabular}{l c c c c c c c}
    \hline\hline
     & \multicolumn{3}{c}{UO$_2$(CO$_3$)$_3^{5-}$} & \multicolumn{3}{c}{NpO$_2$(CO$_3$)$_3^{4-}$} \\
    \cline{2-4}\cline{5-7}
    Method & Curie & LR & total & Curie & LR & total \\
    \hline
    Experiment & -- & -- & $-$62.4 & -- & -- &  $-$93.5 \\
    \hline
    CAS(1,6)\cite{Gendron.Autschbach:Ligand.2016} & $-$82.0 & +24.2 & $-$57.8 & $-$86.5 & +10.8 & $-$75.7 \\
    CAS(7,10)\cite{Gendron.Autschbach:Ligand.2016} & $-$89.6 & +36.4 & $-$53.2 & $-$92.4 & +31.7 & $-$60.7 \\
    RAS[9,100]\cite{Gendron.Autschbach:Ligand.2016} & $-$99.0 & +21.9 & $-$77.1 & $-$110.2 & +17.7 & $-$92.5 \\
    RAS[12,100]\cite{Gendron.Autschbach:Ligand.2016} & $-$99.5 & +24.9 & $-$74.6 & $-$110.4 & +17.8 & $-$92.5 \\
    \hline
    SAP-X2C-PBE0 & $-$38.2 & -- & $-$33.6 & $-$63.6 & -- & $-$64.9 \\
    \hline\hline
  \end{tabular}
\end{table}

The RAS[12,100] total shifts ($-74.6$ and $-92.5$\,ppm) are close to experiment, 
but are the sum of a Curie term and a coupling 
term to the excited states. We though note that the active-space 
dependence of the totals is large especially for the Uranium complex. 
Table~\ref{tab:shifts-carb-budget} analyses the deviation of the shifts in detail. 
The point dipole contribution of our 2c SAP-X2C-PBE0 g-tensor 
agrees with the point-dipole estimates of Ref.~\citenum{Gendron.Sharkas.ea:Calculating.2015}
so the ground-doublet g-tensor is not the cause of the deviation. 
However, the difference lies in the part of the Curie shift that is not point-dipolar.
It is $-$9.1 and $-$28.8\,ppm in SAP-X2C-PBE0 against $-$67.4 and $-$80.8\,ppm 
in the RAS[12,100] calculation (Curie term minus point dipole), 
at metal--carbon distances of 2.96 and 2.88\,\AA. 
Splitting the hyperfine tensor of the carbon into its isotropic and traceless 
parts gives $-$13.7 and $-$21.6\,ppm for the isotropic part and 
$-$24.5 and $-$42.0\,ppm for the traceless part, the latter within 5--7\,ppm 
of the point dipole. 

\begin{table}[b]
  \centering
  \caption{Contributions to the deviation of the shift (calculated minus experimental $^{13}$C shift, ppm) 
  of the carbonates at 273\,K. The individual contributions are given in the order of the rows,
  namely the point dipole of the g-tensor of SAP-X2C-PBE0 is compared against that of 
  Ref~\citenum{Gendron.Autschbach:Ligand.2016}, the hyperfine coupling beyond the point dipole 
  (SAP-X2C-PBE0 against the RAS[12,100] Curie term minus its point dipole), 
  the orbital shielding difference to UO$_2$(CO$_3$)$_3^{4-}$, 
  the LR term to the excited states (absent in this work, entering with opposite sign),
  and the RAS[12,100] result against experiment. This shows that the largest deviation
  is found for the hyperfine coupling beyond the point-dipole contribution.}
  \label{tab:shifts-carb-budget}
  \footnotesize
  \setlength{\tabcolsep}{16pt}
  \begin{tabular}{l c c}
    \hline\hline
    Contribution & U(V) & Np(VI) \\
    \hline
    point dipole of $g$ & +2.9 & $-$5.2 \\
    hyperfine beyond point dipole & +58.4 & +52.0 \\
    orbital difference to U(VI) & +4.6 & $-$1.3 \\
    LR term (absent here) & $-$24.9 & $-$17.9 \\
    RAS[12,100] vs. experiment & $-$12.2 & +1.0 \\
    \hline
    Total deviation & +28.8 & +28.6 \\
    \hline\hline
  \end{tabular}
\end{table}

To better understand the remaining deviation of the 2c SAP-X2C-PBE0 shift
for the actinide carbonyl complexes, we performed a population analysis based on intrinsic atomic orbitals (IAO)
and a natural population analysis (NPA) as outlined in Tables~\ref{tab:carb-pop} and~\ref{tab:carb-spin}.
The IAO $f$ populations of the metals exceed those of the formal 5$f^1$ 
configuration after subtraction of the 4$f$ core, so the yl oxygen atoms (O=An=O) and 
carbonate oxygen atoms donate electrons into the 5$f$ shell. This effect is more 
pronounced for Np. The mean-field spin of the references is only $\langle S\rangle=0.021$ (U) 
and 0.0068 (Np), quenched by spin--orbit coupling from the pseudospin $\tfrac12$. 
The magnetization of the metal is purely $f$. On the ligands it is $p$-type, which is to be expected.
At the carbon atoms the $s$ part is $2\times10^{-5}$ (U) and $7\times10^{-5}$ (Np), 
i.e.\ 2\,\% and 6\,\% of the $p$ part, and the core spin vanishes in the NPA, 
so that valence $s$ spin and core polarization at the carbon nuclei are negligible. 
This is consistent with the small Fermi--contact contribution to the Curie term in 
the wave-function calculations ($+0.1$ and $-10.5$\,ppm, against $-72$ and $-71$\,ppm 
for the orbital and $-27$ and $-29$\,ppm for the spin-dipolar contribution). 
The vectors on the metal-bound oxygen and carbon atoms point in different directions 
and add up to almost zero, outlining that the ligand magnetization is a non-collinear 
spin texture induced by spin--orbit coupling and not a spin density parallel to that of the metal.
However, the individual magnetization of the metal-bound oxygen atoms 
carry 10\,\% (U) and 65\,\% (Np) of that of the central actinide.
As the population analysis characterizes the mean-field reference,
the hyperfine tensor is a derivative with respect to the 
spin orientation, and therefore very sensitive to errors in the latter.
Especially for Np, $^{17}$O NMR spectroscopy is consequently expected to show large 
structural effects.

\begin{table}[htbp]
  \centering
  \caption{Charges and populations of the two-component references of the carbonates
  UO$_2$(CO$_3$)$_3^{5-}$ and NpO$_2$(CO$_3$)$_3^{4-}$.
  IAO and natural charges $q$ and IAO populations per angular momentum
  (mean value over equivalent atoms, the metal $f$ includes the $4f$ core with 14 electrons).}
  \label{tab:carb-pop}
  \footnotesize
  \setlength{\tabcolsep}{7pt}
  \begin{tabular}{l c c c c c c}
    \hline
    Atoms & $q$(IAO) & $q$(NPA) & $n_s$ & $n_p$ & $n_d$ & $n_f$ \\
    \hline
    \multicolumn{7}{c}{UO$_2$(CO$_3$)$_3^{5-}$ } \\
    \hline
     M & +2.94 & +1.47 & 12.33 & 29.85 & 30.00 & 16.88 \\
     O$_{\mathrm{yl}}$ & $-$1.27 & $-$0.94 & 3.90 & 5.37 & 0.00 & 0.00 \\
     O$_{\mathrm{eq}}$ & $-$0.83 & $-$0.79 & 3.71 & 5.12 & 0.00 & 0.00 \\
     O$_{\mathrm{term}}$ & $-$0.79 & $-$0.87 & 3.70 & 5.10 & 0.00 & 0.00 \\
     C & +0.65 & +0.93 & 2.82 & 2.53 & 0.00 & 0.00 \\
    \hline
    \multicolumn{7}{c}{NpO$_2$(CO$_3$)$_3^{4-}$ } \\
    \hline
     M & +3.24 & +1.34 & 12.33 & 29.75 & 30.00 & 17.68 \\
     O$_{\mathrm{yl}}$ & $-$1.04 & $-$0.62 & 3.90 & 5.14 & 0.00 & 0.00 \\
     O$_{\mathrm{eq}}$ & $-$0.81 & $-$0.73 & 3.71 & 5.10 & 0.00 & 0.00 \\
     O$_{\mathrm{term}}$ & $-$0.76 & $-$0.83 & 3.69 & 5.07 & 0.00 & 0.00 \\
     C & +0.65 & +0.93 & 2.82 & 2.53 & 0.00 & 0.00 \\
    \hline\hline
  \end{tabular}
\end{table}

\begin{table}[htbp]
  \centering
  \caption{Spin magnetization $|\mathbf m|=2|\langle\mathbf S\rangle|$ of the two-component references
  of the carbonates in the intrinsic atomic orbitals. The mean magnitude per atom, in percent of the metal (M) value,
  its $s$, $p$, $d$, and $f$ parts, and the magnitude of the vector sum over the equivalent atoms divided by their number (net)
  are listed for UO$_2$(CO$_3$)$_3^{5-}$ and NpO$_2$(CO$_3$)$_3^{4-}$.}
  \label{tab:carb-spin}
  \footnotesize
  \setlength{\tabcolsep}{4pt}
  \begin{tabular}{l c c c c c c c}
    \hline\hline
    Atoms & $|\mathbf m|$ & \% of M & $s$ & $p$ & $d$ & $f$ & net \\
    \hline
    \multicolumn{7}{c}{UO$_2$(CO$_3$)$_3^{5-}$ } \\
    \hline
     M & 0.0459 & 100 & 0.0001 & 0.0005 & $<10^{-4}$ & 0.0463 & 0.0459 \\
     O$_{\mathrm{yl}}$ & 0.0016 & 3 & $<10^{-4}$ & 0.0015 & $<10^{-4}$ & $<10^{-4}$ & 0.0016 \\
     O$_{\mathrm{eq}}$ & 0.0048 & 10 & 0.0001 & 0.0047 & $<10^{-4}$ & $<10^{-4}$ & $<10^{-4}$ \\
     O$_{\mathrm{term}}$ & 0.0003 & 1 & $<10^{-4}$ & 0.0003 & $<10^{-4}$ & $<10^{-4}$ & $<10^{-4}$ \\
     C & 0.0009 & 2 & $<10^{-4}$ & 0.0009 & $<10^{-4}$ & $<10^{-4}$ & $<10^{-4}$ \\
    \hline
    \multicolumn{7}{c}{NpO$_2$(CO$_3$)$_3^{4-}$ } \\
    \hline
     M & 0.0151 & 100 & $<10^{-4}$ & 0.0002 & $<10^{-4}$ & 0.0152 & 0.0151 \\
     O$_{\mathrm{yl}}$ & 0.0005 & 3 & $<10^{-4}$ & 0.0005 & $<10^{-4}$ & $<10^{-4}$ & 0.0005 \\
     O$_{\mathrm{eq}}$ & 0.0098 & 65 & 0.0002 & 0.0096 & $<10^{-4}$ & $<10^{-4}$ & 0.0001 \\
     O$_{\mathrm{term}}$ & 0.0002 & 1 & $<10^{-4}$ & 0.0001 & $<10^{-4}$ & $<10^{-4}$ & $<10^{-4}$ \\
     C & 0.0012 & 8 & 0.0001 & 0.0011 & $<10^{-4}$ & $<10^{-4}$ & $<10^{-4}$ \\
    \hline\hline
  \end{tabular}
\end{table}

\section{Summary \& Conclusion}
\label{sec:conclusion}

We have derived a relativistic two-component Kohn–Sham framework 
for paramagnetic NMR shifts, EPR g and hyperfine tensors, and 
magnetizabilities from a magnetized X2C reference. Spin-orientation 
symmetry breaking together with spin–orbit coupling gives rise to 
non-degenerate Goldstone-like modes, which are parameterized by 
an exponential ansatz. The additional coupled-perturbed Kohn–Sham 
equations yield relaxed tangents, which separate the orbital 
and paramagnetic shielding contributions through an orientation-constrained 
shielding. The g and A tensors follow from the same tangents 
in a single calculation, improving on previous schemes that required three 
SCF calculations with orthogonally aligned spin magnetization. 
The antisymmetric Berry term of the mode dynamics gives access to a 
collective-mode correction of the susceptibility. 
A two-component augmented Roothaan–Hall solver, 
together with a stability analysis, 
ensures that the reference is a true minimum, which is a 
prerequisite for well-defined response equations. Population analyses extended to
two components allow the magnetization of the reference to 
be analyzed. The framework is general and, in conjunction with the SAP-X2C 
Hamiltonian, applicable to heavy-element complexes. Tests 
on Ru(III) and actinide complexes illustrate its use and 
its limits. Natural extensions are the application of other density functionals,
as the theory and implementation are already available up to local hybrids,
and the inclusion of excited-state (Van Vleck-type) contributions.

\section*{Supplementary Material}
Supporting Information is available as a ZIP archive with
\begin{itemize}
   \item Optimized structures (xyz format)
   \item Input files (txt ASCII format)
   \item AO data for intrinsic atomic orbitals (ibodef data) to
   form intrinsic bond orbitals and to perform the initial
   guess with a superposition of atomic densities (txt ASCII format)
\end{itemize}

\begin{acknowledgments}
We thank Christof H\"attig (Bochum) for access to his initial work on the
non-relativistic implementation of the ARH solver.
\end{acknowledgments}

\section*{Author Contributions and Declarations}
\noindent \textbf{Christof Holzer}: Conceptualization (equal);
Data curation (equal); Formal analysis (equal); Investigation (equal);
Methodology (equal); Software (equal); Validation (equal);
Visualization (lead); Writing – original draft (equal);
Writing – review \& editing (equal).
\\
\noindent \textbf{Yannick J. Franzke}: Conceptualization (equal);
Data curation (equal); Formal analysis (equal); Investigation (equal);
Methodology (equal); Software (equal); Validation (equal);
Visualization (supporting); Writing – original draft (equal);
Writing – review \& editing (equal).
\medskip

\noindent \textbf{Notes}: IAO definitions for SAP-X2C were constructed with 
the AI tool Claude Opus/Fable based on standard workflows similar to the H\"uckel
vectors, see Refs.~\citenum{Knizia:Intrinsic.2013, Pollak.Weigend:Segmented.2017, 
Franzke.Spiske.ea:Segmented.2020, Franzke.Holzer:Application.2026}.
Further, Claude was used to validate, comment, and leverage code.
The authors take full responsibility for the code and the theoretical work.

\section*{Data Availability Statement}
The data that support the findings of this study are available within
the article and its supplementary material.
TURBOMOLE is available from the webpage \url{https://www.turbomole.org/}
as compiled binaries. The program suite can be obtained free of charge
for students and educational purposes.

\appendix
\allowdisplaybreaks

\section{Augmented Roothaan--Hall Solver in a Two-Component Formalism}

\subsection{Notation and Preliminaries}
In two-component relativistic Hartree--Fock (HF) theory, the electronic energy can be written as
\begin{equation}
	E_{\mathrm{HF}}[\mathbf{D}]
	= \Tr(\mathbf{h} \mathbf{D})
	+\frac{1}{2} \Tr\!\left(\mathbf{D}\cK[\mathbf{D}]\right)
\end{equation}
where the two-electron map ${\cal K}[\mathbf{D}]$ is linear in the full complex spinor
density $\mathbf{D}$ and self-adjoint. Both terms are real. Hence,
\begin{equation}
	\mathbf F[\mathbf D]=\mathbf h+\cK[\mathbf D], \qquad
	\cK[\mathbf D_l-\mathbf D_m]=\mathbf F[\mathbf D_l]-\mathbf F[\mathbf D_m],
\end{equation}
and the density Hessian is constant. The ARH secant relation is
therefore exact for 2c HF---just like in the non-relativistic 1c HF limit.
For Kohn--Sham DFT, it is an approximation and the DFT secants define
a local quasi-Newton approximation, not an exact global Hessian.
Therefore, the equations of 2c ARH are the complex extension of the 1c equations on a
$2\times2$ spin superstructure. The basis contains $n=2n_{\mathrm{bf}}$ spinor functions and $N_{\mathrm{occ}}$
is the number of occupied spinor states (occupation numbers 0 or 1), and every
matrix operation in the solver becomes complex with the given dimension.
Below, we follow the density-matrix ARH construction of H{\o}st \textit{et al.},~\cite{Host2008,Host2008-2}
the trust-region SCF analyses of Th{\o}gersen \textit{et al.},~\cite{Thogersen.Olsen.ea:trust-region.2004, Thogersen.Olsen.ea:trust-region.2005}
and the orthonormal-basis preconditioning analysis of Sa{\l}ek and co-workers.\cite{Salek2007}
Symmetric matrices become Hermitian, antisymmetric matrices become anti-Hermitian,
real orthogonal matrices become unitary, and the real symmetric eigensolver becomes a
complex Hermitian eigensolver in 2c.

For simplicity and robustness, we work in an orthonormal basis.
Let $\mathbf{S}$ be the spatial AO overlap matrix and let $\mathbf{O}$ be any factor with
\begin{equation}
	\mathbf O\mathbf O^{\dagger}=\mathbf S .
\end{equation}
In the absence of a magnetic field $\mathbf{S}$ is real and the overlap is spin-diagonal. Hence,
the two-component transformation is given as
\begin{equation}
	\mathbf O_{2c} =
	\begin{pmatrix}
		\mathbf O & 0 \\
		0 & \mathbf O
	\end{pmatrix}.
\end{equation}
With finite magnetic fields, the overlap matrix becomes complex and so does $\mathbf O_{2c}$.
For an AO density $\mathbf D_{\mathrm{AO}}$, AO Fock matrix $\mathbf
F_{\mathrm{AO}}$, and AO spinor coefficient matrix $\mathbf C_{\mathrm{AO}}$,
the orthonormal representation is
\begin{align}
	\mathbf P &= \mathbf O_{2c}^{\dagger}\mathbf D_{\mathrm{AO}}\mathbf O_{2c}, \\
	\mathbf F &= \mathbf O_{2c}^{-1}\mathbf F_{\mathrm{AO}}\mathbf O_{2c}^{-\dagger}, \\
	\mathbf C &= \mathbf O_{2c}^{\dagger}\mathbf C_{\mathrm{AO}},
\end{align}
with the subscript AO being applied explicitly for clarity.
The inverse transformations are
\begin{align}
	\mathbf D_{\mathrm{AO}} &= \mathbf O_{2c}^{-\dagger}\mathbf P \mathbf O_{2c}^{-1}, \\
	\mathbf C_{\mathrm{AO}} &= \mathbf O_{2c}^{-\dagger}\mathbf C.
\end{align}

The projector and the Fock matrix obey
\begin{equation}
	\mathbf P^{\dagger}=\mathbf P, \qquad \mathbf P^2=\mathbf P, \qquad \Tr \mathbf P=N_{\mathrm{occ}} =N_{\mathrm{el}},
	\qquad \mathbf F^{\dagger}=\mathbf F.
\end{equation}
This basis is fixed for the whole ARH calculation and is the one in which the density
and the Fock records are stored on disk. The inner solver (micro-iterations) does not work
in this basis but in the semicanonical spinor basis $\mathbf C$ of the current expansion
point, where
\begin{equation}
	\mathbf P = \operatorname{diag}(1,\dots,1,0,\dots,0), \qquad
	\mathbf F_{oo}, \mathbf F_{vv} \ \text{diagonal}.
\end{equation}
All rotations inside the occupied space and inside the virtual space are redundant and
do not change $\mathbf{P}$. The non-redundant rotations are the occupied--virtual and the
virtual--occupied ones and the projection onto them is given by
\begin{equation}
	\Pi(\mathbf X)=\mathbf P_v\mathbf X\mathbf P_o+\mathbf P_o\mathbf X\mathbf P_v ,
	\qquad
	\mathbf P_o=\mathbf P ,
	\quad
	\mathbf P_v=\mathbf 1-\mathbf P ,
	\label{eq:pi}
\end{equation}
which for an anti-Hermitian rotation matrix $\mathbf X$ is the same matrix as $\mathbf
X-\mathbf P_o\mathbf X\mathbf P_o-\mathbf P_v\mathbf X\mathbf P_v$.
For a non-redundant rotation in the current occupied--virtual basis
\begin{equation}
	\mathbf X =
	\begin{pmatrix}
		0 & -\boldsymbol\kappa^{\dagger} \\
		\boldsymbol\kappa & 0
	\end{pmatrix},
\end{equation}
the full-matrix norm follows from the inner product
$(\mathbf A,\mathbf B)=\operatorname{Re}\Tr(\mathbf A^{\dagger}\mathbf B)$ as
\begin{equation}
	\lVert \mathbf X\rVert^2 = (\mathbf X,\mathbf X) = 2\Tr(\boldsymbol\kappa^{\dagger}\boldsymbol\kappa).
\end{equation}
The factor of two is applied so that the trust-radius parameters keep the same meaning
in both the 2c formalism and the existing non-relativistic ARH implementation.

The projector at an SCF trial point reads
\begin{equation}
	\mathbf P(\mathbf X)=e^{-\mathbf X}\mathbf P e^{\mathbf X},
\end{equation}
and its first-order change is given as
\begin{equation}
	\delta \mathbf P = [\mathbf P,\mathbf X] = \mathbf P\mathbf X-\mathbf X\mathbf P.
\end{equation}
Overall, the solver propagates the projector but not the spinors. Here, the exponential is formed
from the eigendecomposition of the self-adjoint $\ii\mathbf X$ and the exponentials of its eigenvalues.
After a propagation step, we form $\mathbf P(\mathbf X)$ and fully restore idempotency by the iterative
McWeeny purification,
\begin{equation}
	\mathbf P \leftarrow 3\mathbf P^2-2\mathbf P^3.
\end{equation}
This step is applied for stability to remove a numerical error that finite arithmetic may cause.
For completeness, the orthonormal spinors transform as $\mathbf C(\mathbf X)=e^{-\mathbf X}\mathbf C$.

\subsection{Gradient, Hessian, and ARH Equation}
The first-order energy change follows as $\delta E=\Tr(\mathbf
F\,[\mathbf P,\mathbf X])$, which is real as $\mathbf F$ and $[\mathbf P,\mathbf X]$
are Hermitian, and thus the gradient with respect to the rotation reads
\begin{equation}
	\cG=[\mathbf P,\mathbf F] ,
	\qquad
	\mathcal{G}_{ai}=-F_{ai}.
\end{equation}
The gradient norm measures the distance from a stationary point in the SCF procedure.

Differentiating further gives two distinct contributions $\boldsymbol\sigma$.\cite{Host2008, Ziolkowski.Weijo.ea:efficient.2008}
The first contribution is the Roothaan--Hall part and keeps the Fock matrix fixed.
It is available and evaluated in closed form,
\begin{equation}
	\boldsymbol\sigma_{\mathrm{RH}}(\mathbf X)
	= \mathbf F_{vv}\mathbf X-\mathbf X\mathbf F_{oo} ,
\end{equation}
written for the occupied--virtual block only. In full matrix form, it is the
anticommutator $\{\mathbf F_{vv}-\mathbf F_{oo},\mathbf X\}$. In the semicanonical basis,
both blocks are diagonal and the action is the trivial element-wise scaling by
$\epsilon_a-\epsilon_i$, where $\epsilon_p=F_{pp}$ are the spinor energies in
the given basis. This corresponds to the preconditioner of Sa{\l}ek and co-workers.\cite{Salek2007}
The second contribution describes the augmentation and the response of the Fock matrix
due to the change of the density, $\delta\mathbf F=\cK[\delta\mathbf P]$. It is not computed
directly but it is approximated based on the history. That is, every previous macro-iteration
supplies one exact pair $(\mathbf P_l,\mathbf F_l)$, and their differences to the current
expansion point,
\begin{equation}
	\Delta\mathbf P_l=\mathbf P_l-\mathbf P ,
	\qquad
	\Delta\mathbf F_l=\mathbf F_l-\mathbf F ,
\end{equation}
satisfy the secant relation $\cK[\Delta\mathbf P_l]=\Delta\mathbf F_l$ exactly
whenever the energy is quadratic in the density (i.e.\ for HF, while it is approximate for DFT).
Note that $\mathbf{P}$ without an index is the projector of the current expansion point,
i.e.\ the last accepted density, and $\mathbf P_l$ are matrices from the history.
The response of a trial rotation is thus obtained by expanding its density change in
the already stored differences,
\begin{align}
	\delta\mathbf P\;\approx\; & \sum_l c_l\,\Delta\mathbf P_l,\\
	\sum_m T_{lm}c_m
	= & \Tr(\Delta\mathbf P_l^{\dagger}\delta\mathbf P), \\
	T_{lm} = & \Tr(\Delta\mathbf P_l^{\dagger}\Delta\mathbf P_m) ,
\end{align}
where the traces of products of two Hermitian matrices are real,
and contracting the same coefficients with the stored Fock differences.
The given response contribution to $\boldsymbol{\sigma}$, i.e\ the augmentation, then reads
\begin{equation}
	\boldsymbol\sigma_{\mathrm{aug}}(\mathbf X)
	={\Pi}\Big(\Big[\mathbf P,\sum_l c_l\,\Delta\mathbf F_l\Big]\Big) ,
\end{equation}
with ${\Pi}$ indicating the projection of Eq.~\eqref{eq:pi} acting on the
matrix in parentheses. The metric $\mathbf T$ is
symmetric positive semidefinite and is inverted with a regularization, as
stored points become linearly dependent when the calculation converges.

Finally, the step is defined from the solution of the level-shifted Newton equation
\begin{equation}
	(\mathbf H-\mu\mathbf 1)\,\mathbf X=-\cG , \qquad \mathbf{H}\mathbf X=\boldsymbol\sigma_{\mathrm{RH}}(\mathbf X)+\boldsymbol\sigma_{\mathrm{aug}}(\mathbf X)
	\label{eq:newton}
\end{equation}
restricted to the occupied--virtual rotations. The shift $\mu$ is negative and
is what turns a Newton step, which converges to the nearest stationary point
of any character, into a descent step. This equation is solved iteratively.
Here, each micro-iteration costs one application of $\mathbf H$, which is 
a trivial element-wise scaling plus one contraction against the history but
no additional Fock build at all. The equation can be solved with a complex 2c version of the
conjugate residual with optimal trial vectors (CROP) solver~\cite{ Ziolkowski.Weijo.ea:efficient.2008}
or the trust-region augmented-Hessian (TRAH) solver.\cite{Thogersen.Olsen.ea:trust-region.2004,
	Thogersen.Olsen.ea:trust-region.2005, Helmich-Paris:trust-region.2021}
Generally, TRAH needs more memory and storage but it is often the slightly more stable
of the two on a flat surface with several nearby solutions.
The preconditioning for the solvers in the semicanonical basis reads
\begin{equation}
	d_{ai}(\mu)=\epsilon_a-\epsilon_i-\mu ,
\end{equation}
and is simply applied as a division. To avoid stability issues, we do not scale
components with a negative denominator or a denominator smaller than $10^{-8}$,
as this keeps the applied preconditioner positive definite.

\subsection{History and Records}
Every macro-iteration writes one record or history point consisting of the Fock matrix
and the density. The complete Fock matrix is stored on disk, as the augmentation contracts
$\Delta\mathbf F_l$ directly. However, for the density only the occupied
spinors $\mathbf C^{(l)}$ of that point are written. The projector follows as $\mathbf
P_l=\mathbf C^{(l)}\mathbf C^{(l)\dagger}$. We stress that the augmentation never needs
$\mathbf P_l$ itself. Only its projections onto the occupied and virtual spaces
of the current expansion point are formed once per macro-iteration,
\begin{equation}
	\mathbf A_l=\mathbf C_v^{\dagger}\mathbf C^{(l)} ,
	\qquad
	\mathbf B_l=\mathbf C_o^{\dagger}\mathbf C^{(l)} .
\end{equation}
As usual, only the independent part of a matrix is written.

The differences $\Delta\mathbf P_l=\mathbf P_l-\mathbf P$ are taken against the
current expansion point, so a change of expansion point changes every one of them
and the metric $\mathbf T$. Recomputing $T_{lm}$ from the records is, however, not necessary.
Instead, the Gram matrix of the stored densities is accumulated
\begin{equation}
	G_{lm}=\Tr(\mathbf P_l^{\dagger}\mathbf P_m)
	=\lVert\mathbf C^{(l)\dagger}\mathbf C^{(m)}\rVert_F^{2} .
\end{equation}
Here, one new row and column is added per record and never touched again, and the metric of
the differences against the record $e$ of the expansion point reads
\begin{equation}
	T_{lm}=G_{lm}-G_{le}-G_{em}+G_{ee} ,
	\label{eq:gram}
\end{equation}
which only involves small symmetric matrices. In the final stages of SCF convergence,
the stored points become almost identical. Thus, the $\mathbf T$
of Eq.~\eqref{eq:gram} becomes numerically singular.
Here, the inverse is taken in the subspace of eigenvalues above a threshold so that the
directions below it are discarded. This means that (almost) redundant records are projected
out to ensure robust convergence.

\subsection{Step Control, Trust Radius, and Judgment}
The step is controlled by the level shift, the trust radius, as well as the judgment
of the step. Here, the level shift was already described for the non-relativistic trust-region
SCF method.\cite{Thogersen.Olsen.ea:trust-region.2004, Thogersen.Olsen.ea:trust-region.2005}
A 2c approach requires a small change for stability. That is, the shift is released along a ladder.
We build this ladder based on the gap of the current expansion point
\begin{equation}
	\Delta=\min_{ai}\,(\epsilon_a-\epsilon_i)
	=\epsilon_{\mathrm{LUMO}}-\epsilon_{\mathrm{HOMO}} ,
\end{equation}
with LUMO and HOMO denoting the lowest unoccupied and the highest occupied molecular orbital
or spinor, respectively. In the first macro-iteration, the shift is $-0.05/\max(\Delta,0.025)$
for a positive gap and $1.1\min(\Delta,-1)$ for a negative one. In subsequent macro-iterations,
we multiply it by $0.75$ and by $0.8$ in the final stages. A rejected step (see below) doubles
it or even resets it again when it had already been released.

The norm of the rotation is bounded by a trust radius based on Ref.~\citenum{Thogersen.Olsen.ea:trust-region.2004}
and in addition the largest rotation angle is bounded, as the step physically performs a spinor
rotation and the 2c SCF procedure is usually more demanding than its non-relativistic limit.
Default values for the trust radius and the largest rotation angle are 0.6 and 0.7\, radians,
respectively. Here, the largest singular value of the stored block is exactly the largest angle by
which $e^{-\mathbf X}$ rotates a spinor. For efficiency, we calculate it with a power iteration on
that block. Results are governed by two rigorous bounds
\begin{equation}
	\max_{ai}|X_{ai}| \;\le\; \sigma_{\max} \;\le\; \lVert\mathbf X\rVert/\sqrt2 ,
	\qquad  \lVert\mathbf X\rVert=\sqrt{(\mathbf X,\mathbf X)}.
\end{equation} 
The trust radius and the bound on the largest rotation angle are later adapted by the ratio of the predicted energy
change and the actually calculated energy change in the judgment.

The trial density is handed to the Fock matrix build, where the energy is recomputed exactly and the step is
subsequently judged by
\begin{equation}
	r=\frac{E_{\mathrm{trial}}-E_{\mathrm{ref}}}
	{\Delta E_{\mathrm{pred}}} \label{eq:arh-ratio},
	\qquad
	\Delta E_{\mathrm{pred}}
	=(\cG,\mathbf X)+\tfrac12(\mathbf X,\mathbf H\mathbf X) ,
\end{equation}
where $E_{\mathrm{ref}}$ is the energy of the last accepted SCF iteration or point. Then,
the step is eventually accepted and the trust radius is updated.\cite{Thogersen.Olsen.ea:trust-region.2004, Thogersen.Olsen.ea:trust-region.2005, Host2008}
In contrast, a rejected step restores the previous density and raises the shift.
Steps with a very small predicted energy change are not judged, as the ratio becomes
unstable. In this case, the step is accepted and the shift is not changed. Further,
the history and records include the rejected steps, as these provide important information
about the secants.

\subsection{Kramers Restriction and Kramers Pairing}
Let $\mathbf J$ represent time-reversal symmetry
\begin{equation}
	\mathbf J= \begin{pmatrix}\mathbf 0&\mathbf 1\\-\mathbf 1&\mathbf 0\end{pmatrix}.
\end{equation}
Then, a matrix $\mathbf{A}$ is time-reversal symmetric if
\begin{equation}
	\mathbf A=\mathbf J\mathbf A^{*}\mathbf J^{\dagger} ,
	\qquad
	\cP_\Theta(\mathbf A)
	=\tfrac12\,(\mathbf A+\mathbf J\mathbf A^{*}\mathbf J^{\dagger}) ,
	\label{eq:treven}
\end{equation}
where $\cP_\Theta$ projects onto that set.
A two-component Kramers-restricted framework imposes this structure on the density.
Here, the density records hold the Kramers-even components only, so the Fock matrix
outside the solver belongs to the Kramers-even part of the density, not to the density the
solver holds. We project the start density and every accepted density onto the
Kramers-even subspace. This corresponds to the set of matrices with $\mathbf P=\mathbf J\mathbf
P^{*}\mathbf J^{\dagger}$.
Therefore, the density matrix is Kramers-even by construction and so is the Fock
matrix. Further, its eigenvalues are degenerate in pairs by Kramers' theorem and the
pairs follow, despite this pairing never being imposed directly.
After every update the Hermiticity residual, the idempotency residual,
the electron count and, where applicable, the Kramers residual are measured.

\subsection{Overview of the Workflow}

One step of the ARH solver consists of a macro-iteration, which builds the Fock matrix from the current density
and updates the density matrix, and micro-iterations to solve the Newton equation to get the step for the density
update. The algorithmic workflow is as follows.
\begin{enumerate}
	\item Set up and form the matrix $\mathbf{O}$ with $\mathbf{O}\mathbf{O}^{\dagger}=\mathbf{S}$ as the
	symmetric square root or alternatively consider the Cholesky factors. Form the 2c matrix $\mathbf{O}_{2c}$ of it. Open the records.
	\item Enter the macro-iterations and form $\mathbf{F} = \mathbf{O}_{2c}^{-1}\mathbf{F}_{\mathrm{AO}}\mathbf{O}_{2c}^{-\dagger}$.
	\item At the expansion point, select the occupied spinors and form the idempotent projector
	$\mathbf{P}$. Further, the semicanonical basis and the orthonormal Fock matrix are built. Only the non-redundant occupied--virtual
	block is needed for the next steps.
	\item Form the non-redundant block of the gradient $\cG=[\mathbf{P},\mathbf{F}]$ with $\mathcal{G}_{ai}=-F_{ai}$
	and check convergence.
	\item Write the current pair $(\mathbf{P},\mathbf{F})$ to the record families and update the density-difference metric
	$T_{lm}=(\Delta \mathbf P_l,\Delta \mathbf P_m)$ with $\Delta \mathbf P_l= \mathbf P_l - \mathbf{P}$.
	\item Solve the ARH equations. The level shift ensures that the diagonal preconditioner is positive definite, and the inner solver (micro-iterations)
	solves $(\mathbf{H}-\mu\mathbf 1)\mathbf{X}=-\cG$ with the CROP or the TRAH solver. Generally,
	these micro-iterations involve preconditioning, level-shifted augmented Hessian action, and the quasi-Newton Fock-response
	part of that action.
	\item In the step control and judgment, the predicted change and the curvature along the step are formed and the step is applied.
	The step is then judged and the trust radius is updated.
	\item Hand back the occupied spinors and return the AO density matrix as
	$\mathbf{D}_{\mathrm{AO}}=\mathbf{O}_{2c}^{-\dagger} \mathbf{P} \mathbf{O}_{2c}^{-1}$.
	Notably, spinor energies are available in every iteration, as the semicanonical basis is rebuilt
	in every iteration. In contrast, the Fock matrix build receives a density matrix, not a set of spinors.
\end{enumerate}

\subsection{Technical Details}
Being a trust-radius or Newton--Raphson method, the ARH solver benefits
from a physically strong initial guess. Therefore, we use a 2c superposition
of atomic densities with the same approximations as in the SCF step.
Partial charges for the atoms are formed with the charge equilibrium model
of the D4 framework.\cite{Wittmann.Gordiy.ea:Extension.2024} The H\"uckel
vectors\cite{Franzke.Holzer:Application.2026} or intrinsic atomic orbitals\cite{Knizia:Intrinsic.2013}
are used as the atomic orbitals, and the orbital basis is projected onto these AOs.
The data for the intrinsic atomic orbitals is given for H--Og as part of the
Supporting Information and the IAOs are the default choice for X2C.

In principle, the method converges to the stationary point with the same occupation pattern
as the start or initial guess. The described rotation does not change the occupation.
A start with a negative gap is re-occupied once by the Aufbau principle prior to the first step.
As finding the correct ground state for open-shell species is still non-trivial in 2c
due to many states being close to each other, we always check for (complex) instabilities
and saddle points after convergence. Here, we probe the converged density $\mathbf P_0$ with
the rotated densities $\mathbf P_0(h\mathbf b)=e^{-h\mathbf b}\mathbf P_0e^{h\mathbf b}$ and
form the approximate Hessian product
\begin{equation}
	\mathbf H\mathbf b\approx\frac1h\bigl[\cG(\mathbf P_0(h\mathbf b))
	-\cG(\mathbf P_0)\bigr] ,
\end{equation}
with both gradients as the non-redundant block in the semicanonical basis $\mathbf C_0$ of
$\mathbf P_0$, which at $\mathbf P_0$ is just the gradient itself. Then,
\begin{equation}
	E(\mathbf P_0(h\mathbf b))-E(\mathbf P_0)=\tfrac12h^{2}(\mathbf b,\mathbf H\mathbf b)+\mathcal O(h^{3}) 
\end{equation}
checks every product with a step size of $h = 0.001$ by default, which was validated against
Ref.~\citenum{Kehry.Franzke.ea:Quasirelativistic.2020}. The trial rotations are the trial
vectors of a Davidson solver~\cite{Davidson1975} and are of the same form as $\mathbf{X}$.
These trial vectors $\mathbf{b}$ are also normalized.
In a Kramers-restricted run, the probes are restricted to the time-reversal-symmetric rotations and a
spin-block mode removes the elements between the two spin blocks. The lowest eigenvalue is obtained
with the Davidson solver with two start vectors, whose Ritz values $\theta$
and residuals $\mathbf r=\mathbf H\mathbf u-\theta\,\mathbf u$ of the Ritz vectors $\mathbf u$
are defined there. For challenging systems, the check is extended
to residuals of up to three Ritz pairs. The density is stable when the residuals
of the followed Ritz pairs are below $10^{-3}$ and the lower bound of Kato~\cite{Kato1949}
and Temple~\cite{Temple1928} on the lowest eigenvalue stays above $-10^{-4}$.
A Ritz value below $-10^{-4}$ indicates a saddle point or an unstable solution. The solver
then steps along its Ritz vector within the same run and converges again, for at most a
given number of analysis rounds. This check is not restricted
to the 2c framework but can be applied analogously in scalar or non-relativistic 1c runs.

To remove linear dependencies of the basis set, the solver can use the canonical
orthonormal basis, which discards overlap eigenvectors below a given threshold. 
A shift of $10^6$ on the discarded space is added to the Fock matrix 
the solver reads. With discarded directions, the AO density is returned 
by the coefficient transformation rather than by the congruence with $\mathbf O_{2c}$.

In finite magnetic fields without spin--orbit coupling and a 
field along the $z$ axis, the solver can be used in a spin-block 
mode using only the $\alpha \alpha$ and
$\beta \beta$ blocks interfaced to a 1c SCF infrastructure. 

\section{Augmented Roothaan--Hall Solver with Fractional Occupations and Smearing}

\subsection{Preliminaries and Central Definitions}
Fermi smearing and simulated annealing are common techniques 
to determine the number of unpaired electrons.
\cite{Rabuck.Scuseria:Improving.1999, Nava.Sierka.ea:Density.2003}
This requires fractional occupations, an entropy term for the free energy, and the
ability to change the occupations within the SCF procedure. That is, the occupations
are determined in an outer step between the macro-iterations. 
The ARH solver then has to supply that step and also judge the combined step. 
This requires a clear distinction
between the generally non-idempotent but self-adjoint weighted density 
$\mathbf D$\cite{ZhangYang2026, ZhangChen2026}
and the idempotent projector $\mathbf P$. The projector in the previous section is
the special case $\mathbf D=\mathbf P$. The other notation is unchanged. So,
spinor indices $p$ and $q$ are general, and in the stored block of rotations $a$
labels the row, the less occupied spinor, and $i$ the column, the more occupied
spinor. For integer occupations these are the virtual and the occupied spinors.
This means that the gradient and the Hessian use occupation differences as weights,
and the redundant rotations are defined within a set of equally occupied spinors and not
within the whole occupied space.

Spinors of equal occupation form a so-called class and the given classes are
denoted $k=1,\dots,K$ in the order of decreasing occupation $n_k$.
For clarity, we denote the occupation of a spinor $p$ by $n_p$, with
$n_p=n_k$ for a spinor of the class $k$. Each class $k$
has the projector $\mathbf P_k=\mathbf C_k\mathbf C_k^{\dagger}$ onto its spinors.
As defined previously, $\mathbf C$ is the matrix of the spinor
coefficients in the fixed orthonormal basis and $\mathbf c_p$ is its column $p$.
The density is a linear combination of idempotent projectors and
the occupations are weights according to
\begin{equation}
	\mathbf P_k^{2}=\mathbf P_k ,
	\quad
	\mathbf P_k\mathbf P_{k'}=\delta_{kk'}\mathbf P_k ,
	\quad
	\sum_k\mathbf P_k=\mathbf 1 ,
	\quad
	\mathbf D=\sum_k n_k\mathbf P_k .
	\label{eq:classes}
\end{equation}
The classes are called the full class at the largest occupation, the virtual class
at zero and the active classes with fractional occupations.
The semicanonical basis is then block-structured by these classes.
In every macro-iteration, $\mathbf D$ is diagonalized and the eigenvectors are grouped
into the classes. Here, the columns are ordered full, active, virtual with the order of
the eigensolver kept within a class. The Fock matrix is diagonalized within each class and
the diagonal elements of $\mathbf F$ in that basis correspond to the spinor energies
$\epsilon_p$, and the eigenvalues of $\mathbf D$ are the occupations.
The occupations are described by the vector $\mathbf n=(n_p)$ and by the diagonal matrix
$\mathbf N=\operatorname{diag}(\mathbf n)$. The spinors are the class-blocked semicanonical
spinors and are therefore also natural spinors of $\mathbf D$. $\mathbf C_k$ in Eq.~(\ref{eq:classes})
collects the columns of class $k$, and the density reads
\begin{equation}
	\mathbf D=\mathbf C\,\mathbf N\,\mathbf C^{\dagger} ,
	\qquad
	\Tr\mathbf D=\sum_p n_p=N_{\mathrm{el}} .
	\label{eq:dw}
\end{equation}
The energy is the ensemble functional $E[\mathbf D]$ that
the Fock builders form for fractional occupations, and its derivative
is one Fock matrix,
\begin{equation}
	\delta E= \Tr(\mathbf F\,\delta\mathbf D) ,
	\qquad
	\mathbf F=\mathbf F[\mathbf D] ,
\end{equation}
shared by all spinors. In Hartree--Fock theory, $E$ is again quadratic in
$\mathbf D$, while the exchange--correlation part of Kohn--Sham DFT is a local
approximation.

Non-redundant rotations are those between two classes, 
and the projection onto them is
\begin{equation}
	{\Pi}(\mathbf X)=\mathbf X-\sum_k\mathbf P_k\mathbf X\mathbf P_k.
	\label{eq:pi-fermi}
\end{equation}
With fractional occupations, the classes are taken from the spectrum of $\mathbf D$,
i.e.\ eigenvalues within $10^{-8}$ of each other form a class. The occupation cutoff
of the smearing is $10^{-6}$, an occupation below it is set to zero, so the grouping
drops no electron. Crucially, the grouping only decides which rotations are redundant,
while $\mathbf D$ and every weight use the exact occupations. Spinors are changed by rotations defined as
\begin{equation}
	\mathbf D(\mathbf X)=e^{-\mathbf X}\,\mathbf D\,e^{\mathbf X}
	=\sum_k n_k\,e^{-\mathbf X}\mathbf P_k e^{\mathbf X} ,
	\quad
	\delta\mathbf D=[\mathbf D,\mathbf X]+\mathcal O(\mathbf X^{2}) .
	\label{eq:param-fermi}
\end{equation}
A unitary transformation conserves the spectrum. Therefore, the occupation vector is
also a constant of the rotation. In other words, no step can change the occupations, so
they are changed only by the separate occupation step. In the natural spinors of
$\mathbf D$, the first-order change reads
\begin{equation}
	[\mathbf D,\mathbf X]_{pq}=(n_p-n_q)\,X_{pq} ,
\end{equation}
and vanishes within a class.
The non-redundant rotations are stored as a rectangular block $\mathbf Y$,
where the rows are the active and virtual spinors and the columns are the full and
active spinors. This means that the active spinors are part of both the rows
and columns. Thus, the block describes every pair twice where both spinors
are active, and pairs within one class. These are removed by a so-called mask
that keeps exactly the pairs of two different classes whose row spinor is
the less occupied one. We define the positive occupation difference of every
stored pair that the mask keeps as
\begin{equation}
	\nu_{ai}=n_i-n_a .
	\label{eq:nu}
\end{equation}
For integer occupations, this equals one. The full rotation then reads
\begin{equation}
	\mathbf X=\mathbf Z-\mathbf Z^{\dagger} ,
	\qquad
	\mathbf Z=\mathbf C_{\mathrm{row}}\mathbf Y\mathbf C_{\mathrm{col}}^{\dagger} ,
\end{equation}
and the inner product of two rotations, denoted $1$ and $2$, is given as
$2\operatorname{Re}\Tr(\mathbf Y_1^{\dagger}\mathbf Y_2)$, which counts every
pair only once. Applying ${\Pi}$ in this basis thus just corresponds
to applying the mask.

Finally, the converged state is a stationary point of the free energy
\begin{equation}
	\Omega(\mathbf C,\mathbf n)=E(\mathbf C,\mathbf n)-TS(\mathbf n)
\end{equation}
over spinors and occupations at a given number of electrons, with
\begin{equation}
	n_p=f\bigl((\epsilon_p-\eF)/\theta\bigr) ,
	\qquad
	\sum_p n_p=N_{\mathrm{el}} .
\end{equation}
Here, $\eF$ is the Fermi level. $\theta$ is the smearing width, and $f$ is the
smearing function, $f(x)=\tfrac12\operatorname{erfc}(x)$ with $\theta=4k_{\text{B}}T/\sqrt\pi$
for the error-function smearing or $f(x)=1/(e^{x}+1)$ with $\theta=k_{\text{B}}T$ for
Fermi--Dirac smearing.

\subsection{Gradient, Hessian, and ARH Equation}
The derivation of the ARH working equations then follows the conventional
formulation. With $\delta E= \Tr(\mathbf F[\mathbf D,\mathbf X])=(\cG,\mathbf X)$,
the gradient reads
\begin{equation}
	{\cG}=[\mathbf D,\mathbf F] ,
	\qquad
	\mathcal G_{ai}=(n_a-n_i)\,F_{ai}=-\nu_{ai}\,F_{ai} ,
	\label{eq:grad-fermi}
\end{equation}
and the second-order term of $E(\mathbf X)$ at fixed $\mathbf F$ is
$\tfrac12 \Tr(\mathbf F[[\mathbf D,\mathbf X],\mathbf X])$,
so that its Hessian action is given by
\begin{equation}
	\boldsymbol\sigma_{\mathrm{RH}}(\mathbf X)
	={{\Pi}}\Big(\mathbf F\mathbf X\mathbf D+\mathbf D\mathbf X\mathbf F
	-\tfrac12\{\mathbf D\mathbf F+\mathbf F\mathbf D,\mathbf X\}\Big) ,
	\label{eq:rh-fermi}
\end{equation}
which reduces to the integer expression $\mathbf F_{vv}\mathbf X-\mathbf X\mathbf F_{oo}$ of the preceding section
for $\mathbf D=\mathbf P$. In the class-blocked basis, the diagonal part of the action
maps a stored element $Y_{ai}$ to $ \nu_{ai}\,(\epsilon_a-\epsilon_i)\,Y_{ai}$.
Here, the prefactor is positive when the more occupied spinor $i$ lies below $a$
and this is the preconditioner of Zhang and Chen,\cite{ZhangChen2026} see also
below. Notably, the Hessian action is not diagonal in this basis. Three spinors
$a$, $i$, and $p$ of three different classes couple through the elements $F_{ap}$
and $F_{pi}$, which are non-zero before convergence, as the Fock matrix is only
diagonalized within the classes.

The augmentation is still of the same structure as for integer occupations. 
However, $\mathbf D$ replaces $\mathbf P$, resulting in
\begin{equation}
	\boldsymbol\sigma_{\mathrm{aug}}(\mathbf X)
	=\Pi\Big(\Big[\mathbf D,\sum_l c_l\,\Delta\mathbf F_l\Big]\Big) ,
	\qquad
	\Delta\mathbf D_l=\mathbf D_l-\mathbf D ,
\end{equation}
as the secant relation requires $\mathbf F$ to be (nearly) linear in the density
but not idempotency itself.\cite{Host2008} Again, we use the labels $l$ and $m$
for the records. This augmentation corresponds to the total-density form of the
ARH Hessian of Zhang and Chen.\cite{ZhangChen2026}
Note that $\boldsymbol\sigma_{\mathrm{RH}}$ is still self-adjoint, 
while the augmented Hessian remains non-self-adjoint.

\subsection{Solving the ARH Equation}
The level-shifted Newton equation is of the same form as for integer occupations
and is solved the same way with the CROP or TRAH algorithms. Here, the shift acts
uniformly on the rotation, so a large shift damps the rotations between spinors
of nearly equal occupation, $Y_{ai}\approx\nu_{ai}F_{ai}/|\mu|$, which then converge
once the shift is released. The preconditioner reads
\begin{equation}
	d_{ai}(\mu)=\nu_{ai}(\epsilon_a-\epsilon_i)-\mu ,
\end{equation}
again leaving non-positive or almost-zero denominators unscaled for numerical
stability. Alternatively, the shift can be weighted with the occupation difference,
which results in the spinor-energy level shift $\nu_{ai}(\epsilon_a-\epsilon_i-\mu)$
on every pair. Furthermore, three important distinctions from the
integer-occupation ARH algorithm apply.

First, a record $l$ now holds the weighted spinors $\Ct_l$, formed from its
natural spinors $\mathbf C^{(l)}$ and its occupations $\mathbf N_l$,
\begin{equation}
	\Ct_l=\mathbf C^{(l)}\,\mathbf N_l^{1/2} ,
	\qquad
	\mathbf D_l=\Ct_l\Ct_l^{\dagger} ,
\end{equation}
over the columns with an occupation above the given threshold of $10^{-8}$.
The number of columns is stored per record, as it may differ between the
records of a smeared run.
The Gram matrix $G_{lm}=\lVert\Ct_l^{\dagger}\Ct_m\rVert_F^2$ as well as
the projections $\mathbf A_l$ and $\mathbf B_l$ are then formed with $\Ct_l$.
Notably, a record stays valid even when the occupations change, as
$\mathbf F_l=\mathbf F[\mathbf D_l]$ regardless of how $\mathbf D_l$ was obtained.
The history of these records across all occupation changes is later used for the
occupation step.

Second, the level-shift ladder uses a generalized gap
\begin{equation}
	\Delta=\min_{k<k'}\,(n_k-n_{k'})\Big[\lambda_{\min}\big(\mathbf P_{k'}\mathbf F\mathbf P_{k'}\big)
	-\lambda_{\max}\big(\mathbf P_k\mathbf F\mathbf P_k\big)\Big] ,
\end{equation}
with the classes ordered by decreasing occupation. $\lambda_{\min}$ and
$\lambda_{\max}$ denote the lowest and the highest eigenvalue of the Fock matrix
projected onto the space of a class. The trust radius and the angle bound
are applied similarly to the conventional ARH algorithm. Here, the trust radius
on $\mathbf X$ is a conservative bound on the density change due to
$\lVert\delta\mathbf D\rVert^2=2\sum_{ai}\nu_{ai}^2|Y_{ai}|^2\le\lVert\mathbf X\rVert^2$.
The power iteration for the largest angle runs on the expanded rotation, as the singular
values of the masked block differ from those of $\mathbf X$.

Third, the density update $e^{-\mathbf X}\mathbf D e^{\mathbf X}$ uses the same
exponential but without McWeeny purification if an active class is present.
A McWeeny purification would drive every occupation to zero or one and consequently
destroy the spectrum the rotation conserves.
In a Kramers-restricted approach, Kramers partners have equal energies and
equal occupations. Therefore, $\mathbf D$ is explicitly time-reversal symmetric,
its classes have even dimension, and the projection $\cP_\Theta$ is applied as
outlined in the last section.

\subsection{Fermi Smearing and the Occupation Step}
The ARH solver itself does not optimize the occupations. These are determined
in an outer step between the macro-iterations, but the solver supplies that step
with the information it requires and then judges the combined step.
According to Janak's theorem,\cite{Janak1978} $\partial E/\partial n_p=\epsilon_p$
holds at fixed spinors. Therefore, the free energy is stationary in the occupations
exactly when every $n_p$ is the smearing function of its own $\epsilon_p$ at a
common Fermi level. The simplest, plain occupation update uses this condition directly.
That is, the new occupations are the smearing function evaluated at the spinor energies
of the last Fock matrix. Here, the spinor energies are kept fixed, and $\eF$ is
determined such that the occupations sum to $N_{\mathrm{el}}$.
This is the occupation update of a conventional SCF procedure with Fermi smearing.
As the spinor energies themselves depend on the occupations, this plain occupation
update is a fixed-point map $\mathbf n\mapsto f(\boldsymbol\epsilon(\mathbf n))$
with the gain
\begin{equation}
	f_{\text{gain}}\approx\frac{1}{4k_BT}\,\frac{\partial\epsilon_p}{\partial n_p} ,
\end{equation}
that is $263\,E_{\text{h}}^{-1}$ times the response at 300\,K and $1580\,E_{\text{h}}^{-1}$
times the response at 50\,K. When $f_{\text{gain}}$ exceeds one, which at the low final
temperatures of an annealing run already happens for a response below a millihartree per
electron, the plain occupation update seriously fails and the occupations oscillate between
two states.

Therefore, the occupation step further uses the response of the spinor energies to the
occupations, which the history supplies. An occupation change is a density change
$\mathbf C\,\delta\mathbf N\,\mathbf C^{\dagger}$, and the augmentation answers it
the same way as it answers a rotation. That is,
\begin{align}
	J_{pq} = & \frac{\partial\epsilon_p}{\partial n_q}
	=\sum_{lm}\Delta F_{l,p}\,\big(\mathbf T^{+}\big)_{lm}\,\Delta D_{m,q} , \\
	\Delta F_{l,p} = & \mathbf c_p^{\dagger}\Delta\mathbf F_l\,\mathbf c_p , \\
	\Delta D_{m,q} = & \lVert\Ct_m^{\dagger}\mathbf c_q\rVert^{2}-n_q ,
\end{align}
with $\mathbf T^{+}$ being the regularized inverse the macro-iteration already holds.
$\mathbf J$ is formed over a so-called window of spinors whose occupation a Fermi level
can move, namely the fractionally occupied spinors, those within eight smearing widths
of the Fermi level, their neighbors, and those the previous step moved. It needs at
least three records. A Newton step at fixed spinors then makes the model free
energy
\begin{equation}
	\Omega_{\mathrm{mod}}(\mathbf n)=\sum_p\epsilon_p\,\delta n_p
	+\tfrac12\sum_{pq}J_{pq}\,\delta n_p\,\delta n_q-TS(\mathbf n)
\end{equation}
stationary over the window at fixed electron count. The response that
makes the plain occupation update fail enters the Newton matrix instead of
the iteration. With $\mathbf J=0$ the plain occupation update is recovered.
The Newton step is safeguarded in four ways. First, a line search on $\Omega_{\mathrm{mod}}$.
Second, a level shift on the occupation Hessian where the exchange part of $\mathbf J$
outweighs the entropy curvature. Third, a refit of $\eF$ to the electron count
bracketed by the window energies. Fourth, a fallback to the plain occupation update
if the refit misses the count.

Among the plain occupation update, the Newton step, and the unchanged occupations,
the one with the lowest model free energy is taken, and an occupation trust radius
(initially set to 0.3 electrons) bounds the change per macro-iteration.

The macro-iteration $t$ then proceeds as follows. The Fock matrix of $\mathbf D^{(t)}$
is built, the solver judges the previous step as outlined in the next subsection,
solves for a rotation, and returns the natural spinors and spinor energies of its
trial density together with $\mathbf J$. The occupation step yields $\mathbf n^{(t+1)}$,
and the trial density
\begin{equation}
	\mathbf D^{(t+1)}=\mathbf C_{\mathrm{trial}}\,\mathbf N^{(t+1)}\,\mathbf C_{\mathrm{trial}}^{\dagger}
\end{equation}
is handed to the next Fock matrix build, i.e.\ the two-electron part (Coulomb,
exchange, and exchange--correlation).
As the returned spinors are its natural spinors, the record of the new point is the
column scaling $\Ct=\mathbf C_{\mathrm{trial}}(\mathbf N^{(t+1)})^{1/2}$, and no further
diagonalization is needed.

\subsection{Judgment of the Occupation Step and Change}
The step is again judged based on the ratio of the actual energy change and the
predicted energy change in Eq.~\eqref{eq:arh-ratio}.
Based on our central definitions, the rotation model includes neither the occupation
step nor the occupation change, but the energy of the occupation change is known to second order.
This allows us to express the predicted energy change as
\begin{equation}
	\Delta E_{\mathrm{pred}}
	=(\cG,\mathbf X)+\tfrac12(\mathbf X,\mathbf H\mathbf X)
	+\sum_p\epsilon_p\,\delta n_p
	+\tfrac12\sum_{pq}J_{pq}\,\delta n_p\,\delta n_q .
\end{equation}
The entropy term, $-T\,\delta S$, is omitted here, as the occupation step prescribes
$\delta\mathbf n$ and nothing models it, so $-T\,\delta S$ cancels between the
predicted and the actual change. Rejections are blocked during occupation changes,
as a smearing step may raise the energy but lower the free energy. Such a rejection
could not be cured either, as a shorter rotation meets the same occupation change
and fails again. The ratio of the actual and predicted change controls both the
trust radius of the rotation and the occupation trust radius.
After the occupations are frozen, rejections apply again.
The entropy term $-TS$ is added to the reported energy only
in the converged iteration. The freezing of the occupations may be blocked until
the occupations are converged and the temperature has reached its given final value.

\subsection{Overview of the Adapted Workflow}
Compared with the integer solver, the steps of the workflow change as follows.
\begin{enumerate}
	\item Setup and the transformation $\mathbf O_{2c}$ are unchanged.
	\item The expansion point forms the weighted density $\mathbf D$ from the
	weighted spinors, and the class-blocked semicanonical basis replaces the
	conventional semicanonical basis.
	\item The gradient carries the pair weights, $\mathcal G_{ai}=-\nu_{ai}F_{ai}$, on the
	masked block.
	\item The history stores weighted spinors and remains stored after occupation changes.
	\item The ARH equations are solved with the class-coupled Roothaan--Hall action
	using the CROP or TRAH algorithms with the preconditioner $\nu_{ai}(\epsilon_a-\epsilon_i)-\mu$.
	\item The update skips purification as soon as an active class exists, i.e.\ as soon
	as fractional occupations are found.
	\item The solver returns the natural spinors, spinor energies, and the
	response $\mathbf J$ of the trial density. The outer occupation step is
	performed and we rebuild the trial density.
	\item The judgment step adds the first-order and the second-order occupation terms
	to the predicted change and blocks rejections when the occupations move.
	\item The optional stability analysis is carried out with fixed occupations after
	convergence. It analyzes the weighted density. The preconditioner and the mask
	of the redundant rotations within a class act in the class-blocked basis of the
	converged density. Here, the SCF procedure may even have converged to integer
	occupations and the analysis requires the final temperature to be reached.
\end{enumerate}

\subsection{Comparison with Restricted Open-Shell Hartree--Fock}
In the non-relativistic 1c limit, the same shell structure underlies the restricted
open-shell Hartree--Fock (ROHF) solver.\cite{Roothaan:Self-Consistent.1960, ZhangYang2026}
Here, fixed occupations are assumed so that the Fermi smearing contribution and the
occupation step are not needed. The orbitals are partitioned into the closed shells $c$,
the open shells $j=1,\dots,N_o$, and the virtual space $v$. In contrast to the classes above,
every shell $k$ has its own Fock operator $\mathbf F_k$ with Roothaan's coupling coefficients.
Therefore, two open shells of equal occupation but different coupling remain separate shells.
The projector onto a shell $k$ is
$\mathbf P_k=\mathbf C_k\mathbf C_k^{\mathrm T}$, with
\begin{equation}
	\mathbf P_k^2=\mathbf P_k,\quad \mathbf P_k\mathbf P_{k'}=\delta_{kk'}\mathbf P_k,\quad
	\mathbf P_c+\sum_j\mathbf P_j+\mathbf P_v=\mathbf 1 .
	\label{eq:proj}
\end{equation}
Here, every orbital of a shell $k$ carries the occupation number $n_k$, with $n_c=2$
for the closed shells, as the restricted 1c case counts up to two electrons per
orbital instead of one per spinor, and $0<n_j\le 2$ for the open shells ($n_j=2f_j$ with
the fractional occupation $f_j$ of Roothaan's formulation), and $n_v=0$ for the virtual
orbitals. The (weighted) density matrices are available from the occupation weights and
the projection as
\begin{equation}
	\mathbf D_c = 2\,\mathbf P_c,\qquad
	\mathbf D_o^{\,j} = n_j\,\mathbf P_j,\qquad
	\mathbf D = \mathbf D_c+\sum_j\mathbf D_o^{\,j} .
	\label{eq:dens}
\end{equation}
In the sign convention of the present work, the gradient reads
$\cG=\Pi\bigl(\sum_k n_k[\mathbf P_k,\mathbf F_k]\bigr)$, and with one Fock matrix
shared by all shells, $\mathbf F_k=\mathbf F$, it becomes $[\mathbf D,\mathbf F]$ and
the equations of the preceding subsections are recovered. In contrast to the class mode
above, which propagates the weighted density $\mathbf D$, the ROHF solver propagates every
shell projector $\mathbf P_k$ and purifies each of them separately by McWeeny purification
after the update.
The ROHF ARH solver only requires real algebra. So, Hermitian matrices
correspond to symmetric matrices and anti-Hermitian ones to antisymmetric
matrices.

\section{Population Analyses in 2c Framework}
We briefly describe the extension of population analysis methods to
calculate partial charges, bond orders, and spin populations from 1c to 2c.
The overlap is always spin-independent and diagonal in the Pauli superstructure,
$\Sm\otimes I$. Without magnetic fields it is real and symmetric, but complex
Hermitian with London orbitals. The number of electrons and the spin expectation
value are available as
\begin{equation}
  \tr\big(\D\,(\Sm\otimes I)\big) = \tr(\Dn\Sm) = N ,\qquad
  \tr(\Dm{k}\Sm) = 2\langle \hat{S}_k\rangle .
\end{equation}
In a Kramers-restricted framework, the spin population vanishes for a real overlap matrix.
However, in the present work we focus on the spin populations for EPR and pNMR.

\subsection{Mulliken Population Analysis}

The Mulliken population \cite{Mulliken:Electronic.1955} is available from the diagonal element
of $\D(\Sm\otimes I)$. Here, the sum over the two spin functions of one AO $\chi_\mu$ is simply
the diagonal of $\Dn\Sm$. The real population follows as
\begin{align}
  q_\mu = & \re(\Dn\Sm)_{\mu\mu} = (\re\Dn\re\Sm)_{\mu\mu} - (\im\Dn\im\Sm)_{\mu\mu}, \\
  m_{\mu,k} = & \re(\Dm{k}\Sm)_{\mu\mu} .
\end{align}
The latter term with the imaginary contribution is only needed for finite magnetic fields.
That is, partial charges use the particle contribution $\Dn$ and the spin magnetization at each
atom, $\ve m_A = \sum_{\mu\in A}(m_{\mu,x},m_{\mu,y},m_{\mu,z})$, is computed from the three spin
components. In the absence of magnetic fields, the Mulliken population is available with the
one-component formula applied component by component using the real parts of the density matrix only.

\subsection{L\"owdin Population Analysis}
The L\"owdin population method \cite{Lowdin:Non-Orthogonality.1950} follows the same way. Only the
symmetric orthogonalization replaces the $\mathbf D_X \Sm$ terms by $\Sm^{1/2}\mathbf D_X\Sm^{1/2}$,
with $X = ( n$ \text{or} $m,k)$, so that
\begin{align}
  q_\mu = & \re\big(\Sm^{1/2}\Dn\Sm^{1/2}\big)_{\mu\mu}, \\
  m_{\mu,k} = & \re\big(\Sm^{1/2}\Dm{k}\Sm^{1/2}\big)_{\mu\mu}.
\end{align}

\subsection{Natural Population Analysis}
The 1c natural population analysis~\cite{Reed.Weinstock.ea:Natural.1985} relies on the matrix
$\mathbf P = \Sm\D\Sm$ in the renormalized AO basis. In the 2c formalism, the actual workflow is
essentially unchanged but it constitutes a complex Hermitian problem with the 2c density of Eqs.~\eqref{eq:2c-density}
and \eqref{eq:pauli}, and the generalized matrix $\mathbf P = (\Sm\otimes I)\,\D\,(\Sm\otimes I)$.
The weights are still real and describe occupations but the symmetry average
changes in 2c, as the $m_l$ average is not suited with SOC. In contrast, every shell is related
to the total angular momentum,
\begin{equation}
  |l\,j\,m_j\rangle = \sum_{m_l m_s}\langle l\,m_l\,\tfrac12\,m_s|j\,m_j\rangle\,
  |l\,m_l\rangle|\tfrac12\,m_s\rangle ,\quad j = l\pm\tfrac12 ,
\end{equation}
where $m_l$ and $m_s = \pm\frac12$ are the projections of the orbital angular momentum and the spin,
$m_s = +\frac12$ for $\alpha$ and $-\frac12$ for $\beta$, $m_j = m_l + m_s$ is the
projection of $\ve J$, and $\langle l\,m_l\,\frac12\,m_s|j\,m_j\rangle$ are the Clebsch--Gordan
coefficients. The real solid harmonics of the basis are transformed to the complex generalization.
The basis is subsequently reordered so that the average runs over $m_j$ within each $(A,l,j)$ block
separately. These blocks read $(s,\frac12)$, $(p,\frac12)$, $(p,\frac32)$, \ldots, and are degenerate
by $2l$ and $2l+2$. The averaged block is one matrix over the given shells and solved only once.
In the coupled basis, a shell of angular momentum $l > 0$ splits into the subshells $j = l-\frac12$
with $2l$ and $j = l+\frac12$ with $2l+2$ components, an $s$ shell stays one $j = \frac12$ subshell.
Therefore, the natural minimal basis size follows directly from the existing 1c table.

The calculation of the natural atomic spinors (NAS) and the natural charges follow accordingly.
\cite{Reed.Weinstock.ea:Natural.1985}
We stress that the spin population is analyzed in the natural atomic spinors of the full 2c density matrix
and not just the spin density contribution. The spin population of a natural atomic spinor follows from
$\big(\mathbf C^{\mathrm{NAS}\dagger}(\Sm\otimes I)\mathbf M_k(\Sm\otimes I)
\mathbf C^{\mathrm{NAS}}\big)_{rr}$ using the Hermitian spin densities $\mathbf M_k = \frac12(\sigma_k\D + \D\sigma_k)$.

\subsection{Bader Charges and Spin Population}
The quantum theory of atoms in molecules \cite{Bader:Atoms.1990} (QTAIM) is based on the
density and the spin magnetization from real AOs on a grid
\begin{align}
  n(\ve r) = & \sum_{\mu\nu} (\re\Dn)_{\mu\nu}\,\chi_\mu(\ve r)\chi_\nu(\ve r),\\
  m_k(\ve r) = & \sum_{\mu\nu} (\re\Dm{k})_{\mu\nu}\,\chi_\mu(\ve r)\chi_\nu(\ve r),
\end{align}
and its topology is the same, with critical points, attractors and zero-flux surfaces
$\nabla n\cdot\ve e = 0$, as Pilm\'e \textit{et al}.~\cite{Pilme.Renault.ea:QTAIM.2014}
have already shown for two component approaches. Here, we do not consider finite magnetic
fields for the basins and the quantities on a grid. The charge and spin population follows as
\begin{align}
  N_A = & \int_{\Omega_A} n\,d\ve r,\qquad q_A = Z_A - N_A,\\
  \ve m_A = & \int_{\Omega_A}\big(m_x,m_y,m_z\big)(\ve r)\,d\ve r,
\end{align}
which relies on a numerical integration over the basins $\Omega_A$. Here, spin--orbit coupling
and 2c formalism only change the basins through $\re\Dn$.

\subsection{Population Analysis with Intrinsic Atomic Orbitals}
The IAO charges and construction, described by Knizia in Eqs.~(1)--(3) and Appendix~C of Ref.~\citenum{Knizia:Intrinsic.2013},
is generalized to the 2c framework by lifting the scalar minimal reference basis to two spin
blocks. Therefore, the overlap of his Appendix~C becomes $\Sm\otimes I$ and the occupied coefficients
are complex spinor coefficients (adjoint instead of transpose). Further equations then hold
as in the quaternion form of Senjean \textit{et al}.~\cite{Senjean.Sen.ea:Generalization.2021} and
the occupied projector mixes the spin blocks and thus also the IAOs due to SOC.
The population of an atom follows from Knizia's Eq.~(3) using the 2c density $\D$ of
Eq.~\eqref{eq:2c-density} herein. This is summed over the IAOs of both lifted copies of its
reference functions. However, no generalization to strong magnetic fields is given, as the
tabulated IAOs are a poor approximation for this case.

The magnetization projects the spin of the density onto the IAOs of the atom and this results
in the spin magnetization population
\begin{align}
  m_{A,k} = & \sum_i n_i\sum_{r\in A}\re\Big[T_{ri}^*\,
  \big(\mathbf A^\dagger(\Sm\otimes I)\,\sigma_k\,\mathbf C\big)_{ri}\Big], \\
  \mathbf T = & \mathbf A^\dagger(\Sm\otimes I)\,\mathbf C \qquad \mathbf A^\dagger(\Sm\otimes I)\mathbf A = \mathbf{1},
\end{align}
where $r$ denotes the orthonormal IAOs of $A$ and $i$ the occupied spinors.
$\mathbf{A}$ is the matrix of the IAO coefficients and $\mathbf{C}$ that of
the spinor coefficients. $T_{ri}$ is the amplitude of spinor $i$ on the IAO $r$.

\subsection{Mayer Bond Orders and Valences}
The Mayer \cite{Mayer:Charge.1983} or Wiberg bond indices \cite{Wiberg:Application.1968} 
were already derived for a general single-determinant. \cite{Mayer:Bond.2007} The general
spin--orbital formulation is given in Eq.~(43) of Mayer's review, \cite{Mayer:Bond.2007} which
directly holds for spinors and the 2c density matrix. The valences are also given Mayer's review,
Eqs.~(64)--(67). Thus, the Mayer analysis is available with SOC and finite magnetic fields
accordingly.

\section{Note on 2c Geometry Gradients}
In the present work, all structures were optimized with SOC effects in
a 2c formalism. This does not require the XC kernel but higher-order derivatives
of the basis functions on a grid. \cite{Pople.Gill.ea:Kohn—Sham.1992, Baldes.Weigend:Efficient.2013}
So, both the canonical and the Scalmani--Frisch non-collinear XC formalism are applicable.
However, we point out that all variables in the canonical formalism depend on the
components of the spin magnetization vector $m_k$ through the direction $m_k/|m|$
and the GGA or mGGA potential in the literature often omit one derivative from the chain rule.
\cite{Van-Wullen:Spin.2002,Armbruster.Weigend.ea:Self-consistent.2008,
Desmarais.Komorovsky.ea:spin-orbit.2021, Holzer.Franzke.ea:Current.2022,Holzer.Franzke:Local.2022}
See Ref.~\citenum{Peralta.Scuseria.ea:Noncollinear.2007} for details on the chain rule
in 2c. Neglecting this term ultimately means that the SCF is not fully variational and
analytic gradients deviate from the derivative of the energy---especially
when the spin magnetization becomes non-collinear. For instance, the deviation between
numerical gradients and analytical gradients at the local minimum is 1--4 orders of
magnitude larger than usually observed in non-relativistic approaches.

With the direction of the spin magnetization $k$, the non-collinear variables of the
canonical GGA and mGGA approach read
\begin{equation}
\hat m_k = \frac{m_k}{|\mathbf m|}, \quad
\boldsymbol{\nabla}|\mathbf m| = \sum_{k=x,y,z} \hat m_k \nabla m_k, \quad
\tau_m = \sum_{k=x,y,z} \hat m_k \tau_k
\end{equation}
so that their derivatives follow as
\begin{equation}
\frac{\partial \boldsymbol{\nabla}|\mathbf m|}{\partial m_k}
  = \frac{1}{|\mathbf m|}\left(\nabla m_k - \hat m_k \nabla|\mathbf m|\right),
  \quad
\frac{\partial \tau_m}{\partial m_k}
  = \frac{1}{|\mathbf m|}\left(\tau_k - \hat m_k \tau_m\right).
\end{equation}
Therefore, spin-dependent part of the exchange-correlation potential of component $u$
gains the additional term
\begin{equation}
v_u^{\text{XC}} \mathrel{+}= \frac{f}{|\mathbf m|}\left[
  \frac{\partial e_{\text{xc}}}{\partial \boldsymbol{\nabla}|\mathbf m|}\cdot
  \left(\nabla m_u - \hat m_u \boldsymbol{\nabla}|\mathbf m|\right)
  + \frac{\partial e_{\text{xc}}}{\partial \tau_k}
  \left(\tau_u - \hat m_u \tau_k\right)\right]
\end{equation}
with the exchange-correlation energy density ${e_{\text{xc}}}$.
Here, the $\tau$-dependent term arises for mGGAs only and $f$ is the damping factor
described below ($f = 1$ for the full term) to ensure numerical stability.
The current-dependent mGGA approach~\cite{Holzer.Franzke.ea:Current.2022} is modified
analogously by adding the term of the projected paramagnetic current density. Local
hybrid functionals follow accordingly with the projected Laplacian, Hessian, and
gradient of $\tau$ for the calibration function.
The standard XC potential in the literature~\cite{Van-Wullen:Spin.2002,
Armbruster.Weigend.ea:Self-consistent.2008, Desmarais.Komorovsky.ea:spin-orbit.2021, 
Holzer.Franzke.ea:Current.2022, Holzer.Franzke:Local.2022} is the special case $f = 0$.
This additional term, however, deteriorates the SCF convergence behavior.
In detail, some grid points show $|m|$ relative to a large density in the
core region without $|m|$ being small in absolute value.
Thus, we apply the damping function
\begin{equation}
f(x) = \begin{cases} 0 & x \le 1 \\
  10 y^3 - 15 y^4 + 6 y^5,\ y = x - 1 & 1 < x < 2 \\
  1 & x \ge 2 \end{cases}
\end{equation}
to improve the SCF behavior. This that the accuracy of the numerical
vs.\ analytical 2c geometry gradients matches that of the non-relativistic
limit.

\section*{References}
\vspace{-0.5cm}
\bibliography{literature}

\end{document}